\documentclass[aps,showpacs,nobibnotes,twocolumn,
nobalancelastpage]{revtex4}
\usepackage{graphicx}
\usepackage{amsmath}
\usepackage{latexsym}
\usepackage{dcolumn}
\usepackage{bm}
\input{epsf}
\newcommand{\simgt}{\lower.5ex\hbox{$\; \buildrel > \over \sim \;$}}
\newcommand{\simlt}{\lower.5ex\hbox{$\; \buildrel < \over \sim \;$}}
\newcommand{\be}{\begin{equation}}
\newcommand{\ee}{\end{equation}}
\newcommand{\ba}{\begin{eqnarray}}

\newcommand{\bi}{\begin{itemize}}
\newcommand{\ei}{\end{itemize}}

\newcommand{\bfi}{\begin{figure}
\epsfxsize=9cm 
\epsffile}
\newcommand{\efi}{\end{figure}}
\newcommand{\la}{\lesssim}
\newcommand{\mnras}{MNRAS}
\newcommand{\apjl}{ApJ}

\newcommand{\aj}{AJ}

\begin{document}
\title{Observational Tests of Modified Gravity}
\author{Bhuvnesh Jain$^{1}$ \& Pengjie Zhang$^{2,3}$}
\email{bjain@physics.upenn.edu}
\affiliation{
${}^1$Department of Physics and Astronomy, University of Pennsylvania,
  Philadelphia, PA 19104\\
${}^2$Shanghai Astronomical Observatory, Shanghai, China 200030\\
${}^3$Joint Institute for Galaxy and Cosmology (JOINGC) of
SHAO and USTC
}

\begin{abstract}
Modifications of general relativity provide an alternative explanation
to dark energy for the observed acceleration of the universe.  
Modified gravity theories have 
richer observational consequences for large-scale structure than 
conventional dark energy models, in that different observables are not 
described by a single growth factor even in the  linear
regime. We examine the relationships between perturbations
in the metric potentials, density and  velocity fields, and discuss
strategies for measuring them using gravitational lensing,  galaxy
cluster abundances, galaxy clustering/dynamics and the ISW effect. 
We show how a broad class of gravity theories can be tested by
combining these probes. A robust way to interpret observations is by
constraining two key functions: 
the ratio of the two metric potentials, and the ratio of the
Gravitational ``constant'' in the Poisson equation to Newton's
constant. We also discuss quasilinear effects that carry signatures of
gravity, such as through induced three-point correlations.  

Clustering of dark energy can mimic features of modified gravity
theories and thus confuse the search for distinct signatures of such
theories. It can produce pressure perturbations and anisotropic stresses, 
which breaks the equality between the two metric potentials even 
in general relativity. With these two extra degrees of freedom, can a
clustered dark energy model mimic  modified gravity models in all
observational tests? 
We show with specific examples that observational constraints on 
both the metric potentials and density perturbations can in principle 
distinguish modifications of gravity from dark energy models. We compare our
result with other recent studies that have slightly different assumptions
(and apparently contradictory conclusions).    
\end{abstract}
\maketitle

\section{Introduction}
The energy contents of the universe pose an interesting puzzle, 
in that general relativity (GR) plus the Standard Model of particle
physics can only account for about $4\%$ of the energy density inferred from
observations.  By introducing dark matter and dark  
energy, which account for the remaining $96\%$ of the total 
energy budget of the universe,  cosmologists have been able to
account for a wide range of  observations, from  
the overall expansion of the universe to 
the large scale structure of the early and late
universe~\cite{Reviews}. 

The dark matter/dark energy scenario assumes the validity of GR at
galactic and cosmological scales and introduces exotic components of
matter and energy to account for observations. Since 
GR has not been tested independently on these scales, a natural
alternative is that the failures of GR plus  the 
Standard Model of particle physics 
imply a failure  of GR. This possibility, that modifications in
GR at  galactic and cosmological
scales can replace dark matter and/or dark energy,  
has become an area of active research in recent years. 

Attempts have been made to modify GR at galactic~\cite{MOND} or cosmological
scales~\cite{DGP,fR,Sahni2005}.   Modified Newtonian Dynamics (MOND) and its
relativistic version (Tensor-Vector-Scalar, TeVeS) ~\cite{MOND} are
able to replace dark matter at galaxy scales to reproduce the galaxy rotation
curves, which provided the earliest and most direct evidences for the
existence of dark matter. The DGP model~\cite{DGP}, in which
gravity lives in a 5D brane world,  naturally leads to late time 
acceleration of the universe. 
Adding a correction term $f(R)$ to the
Einstein-Hilbert action \cite{fR} also allows late time acceleration of
the universe   to be realized.  

In this paper we will focus on modified gravity (MG) theories that are
designed as an 
alternative to dark energy to produce the present day 
acceleration of the universe. In these models, such as DGP and $f(R)$
models, gravity at late
cosmic times and on large-scales departs from the predictions of
GR.  We will consider the prospects
of distinguishing MG models containing dark matter but no dark energy from 
GR models with dark matter and dark energy. By design, successful MG models
will be indistinguishable from viable DE models against observations of
the expansion history  of the universe. To break this degeneracy,
observations of large-scale structure (LSS) must be used to test 
the growth of perturbations. 

LSS in MG theories can be more complicated to predict, but is also 
richer because different observables like lensing and galaxy
clustering probe independent perturbed variables. This differs from 
conventional DE scenarios where the linear growth factor of the
density field fixes all observables on sufficiently large-scales. One
of the goals of this study is to examine carefully what various LSS observables
measure once the assumption of GR (with smooth DE) is dropped.
 
Structure formation in modified gravity in general differs 
\cite{Yukawa,Skordis06,Dodelson06,DGPLSS,consistencycheck,Koyama06,fRLSS,Zhang06,Bean06,MMG,Uzan06,Caldwell07,Amendola07} 
from that in GR. Theories of LSS
in these modified gravity models are still in their infancy. 
However, perturbative calculations at large scales have shown that it
is promising to connect predictions in these theories with
observations of  LSS. Most studies have focused on probes of a single
growth factor with one or a few observables. In this paper we will
consider a variety of LSS observables that can be measured with high precision
with current or planned surveys. Our emphasis will be on
model-independent constraints of MG enabled by combining different
observables. 

Carrying out robust tests of MG in practice is challenging 
as in the absence of a fundamental theory, the modifications to
gravity are often parameterized by 
free functions, to be fine tuned and fixed by observations. Given
the parameter space available to both DE and MG theories, 
it is unclear how the two classes of theories can be distinguished. 
Kunz and Sapone \cite{Kunz06} presented a rather pessimistic
example. They found that one  can tune a clustered dark energy model to
reproduce observations of 
gravitational lensing  and matter fluctuations in the DGP
model. It is not clear if this conclusion applies to all modified
gravity models and if adding more LSS observables helps to break this
severe degeneracy.  

In this paper, we first discuss ways of parameterizing
modified gravity models and dark energy models. 
\S II presents the definitions and evolution equations for
perturbations in the metric and the energy momentum tensor. We then classify
independent LSS observables based on the perturbations that are probed by
them. \S III is devoted to the use of observational probes of LSS for testing
MG. We consider the four fundamental perturbation variables and 
the observations that can be used to probe them. 
The additional information available in the quasilinear
regime is discussed in the Appendix. 
In \S IV we consider the question of distinguishing MG from DE
scenarios. The specific question we want to answer is: 
given a set of LSS observations,
can a general MG model be mimicked by a DE model? If not, what
LSS observables are required to break the degeneracy? We conclude in \S V.

\section{Perturbation Formalism}
\label{sec:formalism}

By definition, the dark sector (dark matter and dark energy) can only
be inferred from their gravitational consequence.
In general relativity,  gravity is determined by the total
stress-energy tensor of all matter and energy ($G_{\mu\nu}=8\pi
G\ T_{\mu\nu}$).  
Thus  we can  treat dark matter and dark energy
as a single entity, without loss of physical generality
\cite{Ma95,Hu99,Kunz07}. This 
entity has total mean matter density $\bar{\rho}_{\rm GR}$ and equation of
state parameter $w=p_{\rm GR}/\bar{\rho}_{\rm GR}$. However, when
discussing  perturbations in this entity, we may 
separate it into a matter component (dissipationless particles
which can be described as a pressure-less fluid free of anisotropic
stress) and a dark energy component. Throughout this paper, when
we refer to ``smooth'' or ``clustered'' dark energy, we refer to this
dark energy subset of the overall dark sector.

We may consider the Hubble parameter $H(z)$ to be fixed by observations. 
In a dark energy model, $\bar{\rho}_{\rm GR}$ is given by the Friedman
equation of GR:  $\bar{\rho}_{\rm GR}=3H^2/8\pi G$.  The equation of
state parameter is $w=-1-2\dot{H}/3H^2$. 

The corresponding modified gravity model has matter
density $\bar{\rho}_{\rm MG}$ to be determined from its Friedman-like
equation. We will consider MG models dominated by dark matter and
baryons at late times and denote fluid variables such as the density 
with subscript $_{\rm MG}$.

\subsection{Metric and fluid perturbations}

With the smooth variables fixed, we will consider perturbations as
a way of testing the models. 
In the Newtonian gauge, scalar perturbations to the metric
are fully specified by two scalar potentials $\psi$ and $\phi$:
\begin{equation}
ds^2 = -(1+2\psi)\ dt^2 + (1-2\phi)\ a^2(t)\ d{\vec x}^2 
\label{eqn:metric}
\end{equation}
where $a(t)$ is the expansion scale factor. This form for the
perturbed metric is fully general for any metric theory of
gravity, aside from having excluded vector and tensor perturbations
(see \cite{Bertschinger2006} and references therein for justifications). 
Note that $\psi$ corresponds to the Newtonian potential for
the acceleration of particles, and that in General Relativity
$\phi=\psi$ in the absence of anisotropic stresses.  

A metric theory of gravity relates the two potentials above to the
perturbed energy-momentum tensor. We introduce variables to
characterize the density and
velocity perturbations for a fluid, which we will use to
describe matter and dark energy (we will also consider pressure and
anisotropic stress below). 
The density fluctuation $\delta$ is given by
\begin{equation}
\delta({\vec x},t) \equiv \frac{\rho({\vec x},t) -
  {\bar\rho(t)} } {\bar\rho(t)} 
\label{eqn:delta}
\end{equation}
where $\rho({\vec x},t)$ is the density and ${\bar\rho(t)}$ is the cosmic
mean density. The second fluid variable is the divergence of the
peculiar velocity 
\begin{equation}
\theta\equiv\nabla_j T_0^j/(\bar{p}+\bar{\rho})={\vec \nabla} \cdot {\vec v}, 
\end{equation}
where $\vec v$ is the (proper) peculiar velocity.  
Choosing
$\theta$ instead of the vector ${\bf v}$ implies that we have assumed
${\bf v}$ to be irrotational. This approximation is sufficiently
accurate in the linear regime, even for unconventional dark energy
models and minimally coupled modified gravity models.

In principle, observations of large-scale structure can directly 
measure the four perturbed variables introduced above: 
the two scalar potentials $\psi$ and $\phi$, and the density and velocity
perturbations specified by $\delta$ and $\theta$. 
As shown below, these variables are the key to distinguishing 
modified gravity models from dark energy. Each has a scale and
redshift dependence, so it is worth noting which variables and at what
scale and redshift are probed by different observations. It is
convenient to work with the Fourier transforms, such as: 
\begin{equation}
\hat\delta(\vec k,t) = \int d^3 x \ \delta(\vec x,t) \
e^{-i {\vec k} \cdot{\vec x}} 
\label{eqn:FT}
\end{equation}
When we refer to length scale $\lambda$, it corresponds to a
a statistic such as the power spectrum on wavenumber $k=2\pi/\lambda$. 
We will henceforth work exclusively with the Fourier space quantities 
and drop the $\hat{}$ symbol for convenience. 

\subsection{Evolution and constraint equations}

We consider here the fluid equations for DE and MG
scenarios. We work in the  Newtonian gauge 
and follow the formalism and notation of \cite{Ma95}, 
except that we use physical time $t$ instead of conformal time. We are
interested in the evolution of perturbations after decoupling, so we
will neglect radiation and neutrinos as sources of perturbations. 
We will make the approximation of
non-relativistic motions and restrict ourselves to sub-horizon length
scales. One can also self-consistently neglect time
derivatives of the metric potentials in comparison to spatial
gradients. These approximations will be referred to as the quasi-static,
Newtonian regime. We will not consider the evolution of perturbations
on super-horizon length scales; \cite{Hu07} show that differences 
in their evolution may have observable consequences for some MG models
(discussed further under the CMB below). 

\subsubsection{Dark Energy with GR scenario} 

We first consider the DE scenario, assuming GR. 
Using the perturbed field equations of GR to first order gives a set of
constraint and evolution equations.
The evolution of the density and velocity perturbations  
includes gravity and pressure perturbations $\delta p$ as sources. 
In the Newtonian limit they give the familiar 
continuity and Euler equations for a perfect fluid. Keeping all
first order terms, and using the notation $\dot\delta\equiv
d\delta/d t$, gives: 
\begin{eqnarray}
\label{eqn:DE1}
\dot{\delta}_{\rm GR}&=&-(1+w)(\frac{\theta_{\rm GR}}{a}-3\dot{\phi})-3H\frac{\delta
  p}{\rho}+3Hw\delta_{\rm GR}\nonumber \\
&\simeq&-(1+w)\frac{\theta_{\rm GR}}{a}-3H\frac{\delta
  p}{\rho}+3Hw\delta_{\rm GR} \ .
\end{eqnarray}
In the second line we have dropped the $\dot{\phi}$ term as it 
is negligible compared to the other terms in the quasi-static regime. 
The Euler equation is given by
\be
\label{eqn:DE2}
\dot{\theta}_{\rm GR}=-H(1-3w)\theta_{\rm GR}-\frac{\dot{w}}{1+w}\theta_{\rm GR}+\left(\frac{\delta
p/\rho}{1+w} -\sigma+\psi\right)\frac{k^2}{a}   
\ee
We have allowed for anisotropic stress sources  in
the energy momentum tensor, parameterized by the scalar $\sigma$,
which enters the Euler equation. 

Note that the above equations describe the multi-component fluid of
baryons, dark matter and dark energy; the density and velocity
variables for this fluid are subscripted $_{GR}$ above
(these variables will represent
a fluid with no dark energy for MG theories below). 
The metric potential variables are $\phi$ and $\psi$ in either
case. Further, we do not subscript $\delta p$ and $\sigma$ as these
sources occur only in the DE plus GR scenario. 

The linearized 
constraint equation gives the Poisson equation for weak field gravity: 
\begin{eqnarray}
k^2\phi&=& -4\pi G a^2\bar{\rho}_{\rm GR}
\left[\delta_{\rm GR}+3(1+w)Ha\frac{\theta_{\rm GR}}{k^2}\right]\ .
\nonumber \\
&\simeq& -4\pi G a^2\bar{\rho}_{\rm GR} \ \delta_{\rm GR}
\label{eqn:DE3}
\end{eqnarray}
where in the second line we have dropped the $H\theta_{GR}/k^2$ term
as it is negligible for nonrelativistic motions on scales well below
the horizon. 

Non-zero anisotropic stress $\sigma$ leads to inequality between the
two potentials:
\be
\label{eqn:DE4}
k^2(\phi-\psi)=12\pi Ga^2(1+w)\bar{\rho}\ \sigma\ \ .
\ee
It is common to take $\phi=\psi$ for ordinary matter and dark matter; 
however clustered dark energy can have a non-negligible anisotropic
stress. 

Eqns. \ref{eqn:DE1}-\ref{eqn:DE4} fully describe the evolution
of perturbations in DE scenarios in the quasi-static, Newtonian
regime. Next we consider the analogous relations for modified gravity
scenarios. 

\subsubsection{Modified Gravity scenario} 

For minimally coupled gravity models with baryons and cold dark
matter, but without dark energy, we can neglect pressure and anisotropic
stress terms in the evolution equations to get the continuity equation: 
\be
\label{eqn:MG1}
\dot{\delta}_{\rm MG}=-\left(\frac{\theta_{\rm
    MG}}{a}-3\dot{\phi}\right)\simeq -\frac{\theta_{\rm
    MG}}{a} \ , 
\ee
where the second equality follows from the quasi-static
    approximation as for GR. The Euler equation is: 
\be
\label{eqn:MG2}
\dot{\theta}_{\rm MG}=-H\theta_{\rm MG}+\frac{k^2\psi}{a} \ .
\ee
For a generic MG theory, the analog of the constraint 
equations (\ref{eqn:DE3}) and (\ref{eqn:DE4}) can take different forms. 
We will attempt to characterize the general behavior in the weak field
limit for small perturbations (small $\delta$) 
and non-relativistic motions. On sub-horizon scales the field
equations in MG theories can then be significantly simplified. 
We parameterize modifications in gravity by two functions 
$\tilde{G}_{\rm  eff}(k,t)$ and $\eta(k,t)$ to get the analog of the
Poisson equation and a second equation connecting $\phi$ and
$\psi$~\cite{Zhang07}. We first write the generalization of the
Poisson equation in terms of an effective gravitational constant
$G_{\rm eff}$:
\be
\label{eqn:MG3a}
k^2 \phi=-4\pi G_{\rm eff}(k,t)\bar{\rho}_{\rm MG}a^2\delta_{\rm
  MG}\ .
\ee
Note that the potential $\phi$ in the Poisson equation comes from the
  spatial part of the metric, whereas it is the ``Newtonian'' potential $\psi$
  that appears in the Euler equation (it is called the
  Newtonian potential   as its gradient gives the acceleration of
  material particles). Thus in MG, one cannot directly use the Poisson
  equation to eliminate the potential in the Euler equation. 
  A more useful version of the Poisson 
  equation would relate the sum of the potentials,
  which determine lensing, with the mass density. We therefore
  introduce $\tilde G_{\rm eff}$ and write the constraint equations for
  MG as
\be
\label{eqn:MG3}
k^2(\psi+\phi)=-8\pi \tilde{G}_{\rm eff}(k,t)\bar{\rho}_{\rm MG}a^2\delta_{\rm
  MG}
\ee
\be
\label{eqn:MG4}
\phi=\psi \ \eta(k,t)
\ee
where $\tilde{G}_{\rm eff} = G_{\rm eff}(1+\eta^{-1})/2$. 
Note that if one starts in real space
then the corresponding parameters would not be Fourier transforms of
$\eta$ and $\tilde{G}_{\rm eff}$. Thus the Fourier transform of the 
PPN parameter $\gamma\equiv\phi/\psi$, the ratio of the metric
potentials in real space constrained by 
solar system tests, is given by a convolution of $\eta$ and
$\psi$ \footnote{We thank Eric Linder for pointing out the caveat about
  real and Fourier space treatments.}. 
Only if $\eta$ is scale 
independent would it be the Fourier transform of $\gamma$. A similar
reasoning applies to $\tilde{G}_{\rm eff}$ in using the Poisson
equation. We prefer to work in Fourier space because of the ease of
describing perturbations: each Fourier mode evolves independently 
in the large-scale, linear regime. Furthermore, the
equations describing cosmological perturbations in MG theories such as $f(R)$
gravity and DGP are generally expressed in Fourier space. 

The parameter $\tilde{G}_{\rm eff}$ characterizes
deviations in the ($\psi+\phi$)-$\delta$ relation from that in GR. Since the
combination $\psi+\phi$ is directly responsible for gravitational
lensing, 
$\tilde{G}_{\rm eff}$ has a specific physical meaning: it determines the
power of matter 
inhomogeneities to distort light. This is the reason we prefer it over
working with more direct generalization of Newton's constant,
${G}_{\rm eff}$.

The $\tilde{G}_{\rm eff}$-$\eta$ 
parameterization is equivalent to the $Q$-$\eta$ parameterization
independently proposed by \cite{Amendola07}, where $Q$ parameterizes
deviations in Poisson equation (\ref{eqn:DE3}) from GR. For minimally
coupled gravity models, with no dark energy fluctuations, it is also
equivalent to that proposed by \cite{Uzan06}. And
$\eta$ is also equivalent to the parameter $\varpi$
proposed by \cite{Caldwell07}. DGP and $f(R)$ gravity can be described
by our parameterization. So as the 
widely adopted Yukawa potential.  An exception to our approach is
TeVeS as it includes  scalar and vector fields that are coupled to the
growth of scalar  perturbations. 

For a generic metric theory of MG, one would expect that a
Poisson-like equation is valid to leading order in the potentials and
the density perturbation, at least on large scales in 
the linear regime where Fourier modes are uncoupled. 
In this regime, we
expect that since the left-hand side of the field
equations involve curvature, it must have second derivatives of the metric
perturbations, while the right hand side is simply given by the energy
momentum tensor.  On smaller scales, in
general a MG theory may not obey superposition and require
higher order terms and higher derivatives of the potentials. 
Similarly a generic relation between $\phi$ and $\psi$ is likely to
have a linearized 
relation of the form in Eqn. \ref{eqn:MG4}. While it is not
necessary that the leading term be linear in both the
potentials, observational constraints require that it be very close to
linear with $\eta\simeq 1$ on small scales where tests of gravity
exist(see \cite{Will2001} for a review). 

With the linearized equations above, the evolution of either the
density or velocity perturbations can be described by a single second
order differential equation. In the case of MG theories, this equation
is simpler 
as the only source is provided by the Newtonian potential $\psi$. From 
Eqns. \ref{eqn:MG1} and \ref{eqn:MG2} we get, for the linear
solution, $\delta(\vec k,t)\simeq \delta_{initial}(\vec k) D(k,t)$, 
\begin{equation}
\ddot{\delta}+2 H \dot{\delta} + \frac{k^2 \psi}{a}
 = 0 . 
\label{eqn:lingrowth}
\end{equation}
For a given theory, Eqns. \ref{eqn:MG3} and \ref{eqn:MG4} then
allow us to substitute for $\psi$ in terms of $\delta$ to
determine $D(k,t)$, the linear growth factor for the density: 
\begin{equation}
\ddot{D}+2 H \dot{D} - \frac{8 \pi \tilde{G}_{\rm eff}}{(1+\eta)}
\bar{\rho}_{MG}a^2\ D = 0 . 
\label{eqn:lingrowth2}
\end{equation}
We can also use the relations given above to obtain the linear growth
factors for $\theta$ and the potentials from $D$. Note that in general 
the growth factors for the potentials have a different $k$ dependence
than $D$. In the Appendix we give details on the linear and second order
solutions and summarize quasilinear signatures of MG theories. 

\subsection{Power spectra}

Before we turn to large-scale structure observables, we define
the power spectra of the perturbed variables. 
The three-dimensional power spectrum 
of $\delta(k,z)$  for instance is defined as
\begin{equation}
\langle \delta({\vec k}, z) \delta({\vec k'}, z) \rangle = 
(2 \pi)^3 \delta_{\rm D}({\vec k + k'}) P_\delta(k,z) .
\label{eqn:powerdef}
\end{equation}
where we have switched the time variable to redshift $z$. The power spectra of 
perturbations in other quantities are defined analogously. We will denote the 
cross-spectra of two different variables with appropriate subscripts, for
example $P_{\delta\psi}$ denotes the cross-spectrum of the density $\delta$
and the potential $\psi$. 

We write down next the relation between the power spectra of the two 
potentials and the density in DE and MG scenarios. 
From the Poisson equation (\ref{eqn:DE3}) for GR we have 
\begin{equation}
GR: 
P_\phi(k, z) = (4\pi G)^2 a^4 \bar\rho_{\rm GR}^2 \frac{P_{\delta,{\rm GR}}(k, z)}
{k^4} \ , 
\label{eqn:power3}
\end{equation}
where $P_\phi$ is the power spectrum of the potential $\phi$. 
Using the Friedman equation for GR the above equation is often written as
\begin{equation}
GR: 
P_\phi(k, z) = \frac{9}{4} H_0^2 \Omega^2 \frac{P_{\delta,{\rm GR}}(k, z)}
{a^2 k^4} , 
\label{eqn:power3b}
\end{equation}
where $H_0$ is the present day value of the Hubble parameter, and
$\Omega$ is the dimensionless density parameter. 

The Poisson equation (\ref{eqn:MG3}) for MG gives the following
equations for the power spectra of the metric potentials. 
\begin{eqnarray}
&MG:& P_{\psi+\phi}(k, z) =
[8\pi \tilde{G}_{\rm eff}(k,z)]^2 a^4 \bar\rho_{\rm MG}^2 
P_{\delta,{\rm MG}}(k, z)/k^4 \nonumber
\\
&{\rm or, }& \ P_\phi = 
\frac{[8\pi \tilde{G}_{\rm eff}(k,z)]^2}{[1+\eta^{-1}(k, z)]^2} a^4 \bar\rho_{\rm MG}^2 
\frac{P_{\delta,{\rm MG}}(k, z)}{k^4\ }
\label{eqn:power4}
\end{eqnarray}
where we have used Eqn. \ref{eqn:MG4} to get the equation for
$P_\phi$. 

For LSS observables, we will need to the power spectra of $(\psi+\phi)$
for lensing, of $\psi$ for dynamics, and of $\delta$ for tracers of
LSS. We will use Eqns. \ref{eqn:power3}-\ref{eqn:power4} above to
connect them, along with the relations between
the two potentials (Eqn. \ref{eqn:DE4} for GR and Eqn. 
\ref{eqn:MG4} for MG). 
With these relations we can express different observable power
spectra in terms of a single density power spectrum -- for MG this
will involve the functions $\tilde{G}_{\rm eff}(k,z)$ and
$\eta(k,z)$. 

\section{Large-scale structure observations}
\label{sec:LSS}

We will assume that the background expansion rate is determined by a
set of observations: Type Ia supernovae, baryon acoustic oscillation
(BAO) and other probes at low  
redshift and the CMB and nucleosynthesis at high redshift. These
observations measure the luminosity or angular diameter distance 
at a given redshift. The distance measures in a spatially flat
universe are, within factors of
$1+z$, simply the comoving coordinate distance:
\begin{equation}
\chi(z) = \int_0^z \frac{dz'}{H(z')}
\label{eqn:distance}
\end{equation}
Furthermore, BAO can directly measure the Hubble constant at the
redshift of galaxies.

We are interested in the constraints available on perturbed quantities. 
Hence we will consider observational probes of large-scale structure
to constrain modified gravity scenarios. In nearly all cases we will
be interested in scales in the range $1-10^3$ Mpc. The MG theories of
interest must modify gravity on horizon scales of order $10^4$ Mpc; it is
an open question how they transition to GR on very small scales to
satisfy experimental constraints from solar system tests. We will
assume that the MG theories of interest differ from GR over the
observationally accessible scales. 

The most stringent current tests of gravity  
come from laboratory and solar system tests and from binary
pulsar observations -- see \cite{Will2001}
for a review. Interesting probes of gravity on sub-Mpc scales also
exist: galaxy rotation curves, satellite dynamics, strong lensing 
observations of galaxies and clusters, and 
X-ray plus lensing observations of clusters
(e.g. \cite{Bolton2006}). Modifications in gravity can affect the
propagation of gravitational wave. Future gravitational wave
experiments such as LISA can detect gravitational wave from distant
supermassive   black hole pairs in the coalescence phase and thus test
this effect \cite{GW}. We will not consider these tests in this paper. 
We will restrict our attention to large-scale structure on scales 
where theoretical predictions 
can be made using linear or quasilinear perturbation theory. 

\subsection{Connection of observables to perturbation variables}
\label{sec:connection}

In principle, observations of large-scale structure can directly 
measure four fundamental variables that describe the perturbed metric 
and (fluid) energy-momentum tensor: the two scalar potentials $\psi$ 
and $\phi$ that characterize the metric, and the density and velocity
perturbations specified by $\delta$ and $\theta$. 
Next we discuss the prospects for different probes of these
variables. 

\bigskip
{\bf Sum of potentials $\psi+\phi$:} Gravitational lensing in either the
weak or strong lensing regime probes the sum of the metric
potentials. We will consider the weak lensing shear (or equivalently
the lensing convergence)  power spectrum as
the primary statistical discriminator of MG via lensing. 

The spatial components
of the geodesic equation for a photon trajectory $x^\mu(\lambda)$
(where $\lambda$ parameterizes the path) is: 
\begin{equation}
\frac{d^2x^\mu}{d\lambda^2} + \Gamma^\mu_{\rho \sigma}
\frac{dx^\rho}{d\lambda} \frac{dx^\sigma}{d\lambda} = 0
\end{equation}
For the metric of Eqn. \ref{eqn:metric}, this gives
the following relation for the first order perturbation to the 
photon trajectory (generalizing for example 
from Eqn. 7.72 of \cite{Carroll2004}):  
\begin{equation}
\frac{d^2 x^{(1)\mu}}{d\lambda^2} = -q^2 \vec{\nabla}_\perp(\psi+\phi)
\ . 
\end{equation}
where $q$ is the norm of the tangent vector of the unperturbed path. 
This gives the deflection angle formula
\begin{equation}
\alpha_i = -\int \partial_i(\psi+\phi) ds \ , 
\end{equation}
where $s=q\lambda$ is the
path length and $\alpha_i$ is the $i-$th component of the deflection
angle (a two-component vector on the sky). 
Since all lensing observables are obtained by taking derivatives
of the deflection angle, they necessarily depend only on the
combination $\psi+\phi$ (to first order in the potentials). 

For weak lensing tomography we use the shear power spectrum for
two sets of source galaxies with redshift
distributions centered at $z_i$ and $z_j$. Following standard
treatments of weak lensing, this may be derived from the deflection
angle formula to get the shear power spectrum on angular wavenumber
$l$ (\cite{Hu1999}):
\begin{equation}
C_{\gamma_i \gamma_j}(l)
=\int\! d\chi W_i\!(\chi) W_j\!(\chi)
k^{-4} P_{\psi+\phi}\!\left(k=\frac{l}{\chi}, \chi\right),
\label{eqn:shearpower}
\end{equation}
where the weight function $W_i$ is simply
\begin{equation}
W_i \propto \frac{\chi_i - \chi}{\chi_i}
\label{eqn:lensing_weight}
\end{equation}
for source galaxies at a single comoving distance
$\chi_i\equiv\chi(z_i)$ (it can be easily generalized for sources specified by a 
redshift distribution). 
We have assumed a flat background geometry for simplicity; our
results throughout this paper can be generalized to a curved
spatial geometry by replacing $\chi$ in the argument of $W$ 
by the angular diameter distance.  

Note that in the literature the lensing power spectra for GR 
are expressed in terms of the density power spectrum $P_\delta(k)$ 
assuming the standard Poisson and Friedman equations.
Usually anisotropic stress is
neglected so that one can substitute into the above equation the 
relation between the power spectra: 
$P_{\psi+\phi} = 9\ k^{-4} H_0^4\ \Omega^2 P_\delta/a^2$ from Eqn. 
\ref{eqn:power3b}. 
For MG, this substitution breaks down due to the modifications of the
Poisson equation and the Friedman equation. However the correct substitution can
be made in terms of $\tilde{G}_{\rm eff}(k,z)$ using equation
\ref{eqn:power4} and the modified Friedman equation (which depends on
the specific theory). 

Since lensing probes the sum of the metric potentials, with the 
deflection angle formula following from the geodesic equation (which simply 
describes how curvature affects trajectories), it may not by itself test 
the field equations of the gravity theory. However lensing measurements at 
multiple source redshifts are sensitive to the growth of the lensing
potential, which does offer a test of the MG theory. And by combining
lensing with other observables, 
the relation of $P_{\psi+\phi}$ to $P_\delta$ can be tested. 
Recent studies that have examined constraints on MG theories with weak
lensing include \cite{Uzan2001,Heavens2007,Amendola07,Caldwell2007,Acquaviva2007}. 

Another important observable in lensing is galaxy-galaxy lensing, the mean
tangential shear around foreground (lens) galaxies. Its Fourier transform, 
the galaxy-lensing cross-spectrum, depends on $\psi+\phi$ and on
the galaxy number density. It is given by an equation similar to
Eqn. \ref{eqn:shearpower}, with the power spectrum of the lensing potential 
in the integrand replaced by the
three-dimensional cross-power spectrum, and with one of the weight
functions replaced by one representing the foreground galaxy distribution: 
\begin{equation}
C_{g_i \gamma_j}(l)
=\int\! d\chi \frac{W_{gi}\!(\chi) W_{\gamma j}\!(\chi)}{k^2\chi }
P_{g(\psi+\phi)}\!(k=\frac{l}{\chi}, \chi),
\label{eqn:galpower}
\end{equation}
where $W_{gi}$ is the normalized (foreground) galaxy redshift
distribution (e.g. \cite{Hu2004}). 
Galaxy-galaxy lensing has been well measured from the SDSS survey. 
It is a very useful check on galaxy bias, hence it aids the interpretation of 
galaxy clustering measurements (\cite{Seljak2006}) as well. 
\\
{\it Assumptions:} In using weak lensing observations with the above
  formalism,  one assumes that intrinsic correlations are
  negligible or removable (in general these can
  differ for different gravity theories), that the weak lensing 
approximation is valid, and that galaxy properties that affect
photometric redshift determination are not affected by the gravity
theory. 

\bigskip
{\bf Newtonian Potential $\psi$:} 
This can be measured by dynamical probes, typically involving galaxy or
cluster velocity measurements. If gravity is the only force
determining galaxy  accelerations at large scales (as expected), 
we have from Eqn. \ref{eqn:MG2}: 
\be
k^2\psi=\frac{d{(a\theta_g)}}{d t}\ ,
\label{eqn:psi-thetag}
\ee
where $\theta_g\equiv \nabla\cdot{\bf v}_g$. 
On sub-Mpc scales this relation can be used to constrain $\psi$ using
galaxy satellite dynamics and rotation curves (e.g. \cite{Klypin2007}). 
Redshift distortion effects in the galaxy power spectrum probe larger
scales, which we address in more detail here. 

The redshift space power spectrum of
galaxies is a well measured quantity. It 
can be expressed in the large-scale, small angle limit as
(e.g. \cite{Scoccimarro04}) : 
\begin{equation}
P_g^s(k) = \left[P_g(k)+\frac{2u^2}{H}P_{g\theta_g}(k)+\frac{u^4}{H^2}
P_{\theta_g}(k)\right]F\left(\frac{k^2u^2 \sigma^2_v}{H^2(z)}\right)
\label{eqn:zspace}
\end{equation}
where $u=k_{\parallel}/k$ is the cosine of the angle of the ${\bf k}$
vector with
respect to radial direction; $P_g$, $P_{g\theta_g}$, $P_{\theta_g}$ are
the real space galaxy  power
spectra of galaxies, galaxy-$\theta_g$ and $\theta_g$, respectively;
  $\sigma_v$ is the 1D velocity dispersion; and $F(x)$ is a
smoothing function, normalized to unity at $x=0$, determined by the
  velocity probability distribution. The dependence on $u$ enables 
  separate measurements of all three power spectra, though $P_{\theta_g}$
  is the hardest to measure with high precision
  \cite{THX02,Tegmark2004}. 
Furthermore,
  measurements of $P_g^s$ at smaller scales provide information on pairwise
  velocity dispersion $\sigma_v$ \cite{v12}. 

In the linear regime, we can rewrite Eqn. \ref{eqn:psi-thetag} as
\be
k^2\psi=\frac{d(aD_{\theta})/dt}{D_{\theta}}{\theta_g}\ . 
\ee
Here $D_{\theta}$ is the growth factor of $\theta_g$. 
For MG models, $D_{\theta}$ has a simple relation to
$D$, the linear density growth factor: $D_{\theta}\propto a\dot{D}=a\beta
HD$, where $\beta\equiv d\ln D/d\ln a$. 
In the linear regime we have
$\theta_g({\bf k}, t)=\theta_g({\bf k},t_i)D_{\theta}({\bf
  k},t)$. 
Note that the above equation does not require $D_{\theta}$ to be scale
independent, so 
it is applicable to modified gravity models and clustered dark energy
models. Note also that we do not distinguish the growth factor of
$\theta_g$ from that of $\theta$ 
because we only use its time (redshift) derivative, which is expected to
be very similar. Velocity  measurements at multiple redshifts are
required to measure $\psi$ from the above equation,  as described in
\cite{Zhang07b}.   

For clustered DE models, the galaxy $v_g$ is not necessarily equal to
$v$ of the total fluid. 
From the Euler equation (\ref{eqn:DE2}) applied separately to different 
components of the fluid, we can see that the DM and DE velocities
evolve differently since only the latter is affected by pressure perturbations 
in the DE. As a first order approximation,  galaxies and
baryonic gas velocities trace that of the DM. So what one actually
measures is $\theta_g\simeq \theta_{\rm DM}\neq \theta_{\rm DE}\neq \theta$.
This distinction can be relevant for DE models with large perturbations 
on sub-horizon scales if these are not correlated with the matter
fluctuations (i.e. if the DE power spectrum has a different shape from
the matter power spectrum). 

{\it Assumptions/Caveats:} 
The galaxy peculiar velocity only probes $\psi$ where
there are galaxies. So potentially there is a 
bias related to the environment of galaxies. However, since gravity is
a long range force, the potential where galaxies reside is determined 
by matter over a much larger region and thus should be unbiased 
with respect to the overall $\psi$. Galaxies themselves are not
sufficiently massive to contribute to this long range
potential. However, to  obtain $\dot{v}_g$ from limited redshift bins,
one does need to 
parameterize the  redshift dependence of $v_g$.  

The accuracy of the velocity information inferred from the redshift
space galaxy  power spectrum relies on the modeling of the redshift
distortion. The derivation of Eqn. \ref{eqn:zspace} is quite
general -- it can be applied to general
DE or MG models. However, Eqn. \ref{eqn:zspace} does not describe
redshift distortions to percent level accuracy
\cite{Scoccimarro04}.  Nonetheless, with improved modeling of the
correlation function in redshift space \cite{Tinker07} the associated
systematic errors in velocity (and $\psi$) measurements can be reduced.  

\bigskip
{\bf Density contrast $\delta$: } 
The clustering of galaxies is one of the earliest measures of large-scale
structure, and its measurements have advanced over the last three
decades.  The galaxy power spectrum $P_{g}$ is the simplest
statistical measure of correlations in the galaxy number
density. Several other probes of large-scale structure also probe the
density field: clustering of the Lyman-alpha forest, clustering of quasars and
galaxy clusters, the abundance of galaxy clusters, and (in the future)
21-cm emission measurements of the high-redshift universe. 

However, given a measured galaxy power spectrum, the
power spectrum $P_\delta$ of the underlying mass density 
$\delta$ may differ due to galaxy bias. 
Further the galaxy-density relation may be
non-local and vary slightly 
in different gravity theories due to differences in the 
tidal field that influence collapsed objects such as galaxy halos. 
We will restrict ourselves to large scales ($k \ll k_{\rm
nl}$, the nonlinear wavenumber) 
where bias is scale independent in simple models of galaxy
formation. This allows us to infer the mass power spectrum from the
galaxy power spectrum without detailed modeling of their relation,
because it is possible to fit for the bias directly from the data.  
We discuss below the caveats to this assumption for clustered dark
energy.

The galaxy density in
three-dimensional space may be expressed in terms of the 
density and bias parameters $b_1$ and $b_2$ as
\begin{equation}
\delta_g \equiv \frac{\delta n_g}{n_g} = b_1 \delta + \frac{b_2}{2} \delta^2 .
\label{eqn:bias}
\end{equation}
This expansion is useful for small values of $\delta$; it can be used
in a perturbative expansion to explore what measurements are
sufficient to measure the bias parameters $b_1$, $b_1$ as well as 
$\delta$ (see \cite{Bernardeau2002} for details on the bias
formalism). Eqn. \ref{eqn:zspace} above 
shows how the three-dimensional galaxy power spectrum $P_g$ can be obtained
from redshift space measurements. 
A second way of measuring $P_g$ is from 
imaging data with photometric redshifts. This provides 
measurements of the angular power spectrum of galaxies, which is a
projection of the three-dimensional galaxy power spectrum 
\begin{equation}
C_{g}(l)
=\int\! d\chi \frac{W_g^2\!(\chi)}{\chi^2}
P_{g}\!\left(k=\frac{l}{\chi}, \chi\right),
\label{eqn:galaxypower}
\end{equation}
where $W_g$ is the normalized redshift distribution of galaxies
included in the sample. With good photo-z's it is a narrow range with
width of order 0.1 in redshift, so that many such angular spectra can
be measured at different mean redshifts from a survey 
(e.g. \cite{SDSSAngular}). 

\subsubsection{Galaxy bias with clustered dark energy}
In clustered dark energy models it is not a priori clear whether the
galaxy overdensity is related to the matter overdensity $\delta_m$ 
or to the total fluid overdensity $\delta_{\rm GR}$. We argue below
that at least for some galaxy populations, $\delta_g$ is directly 
related to $\delta_m$, even
though the evolution of the matter density responds to the full
gravitational potential (which receives contributions from dark
energy clustering as well). 

One way to see this is to consider the centers of mass of galaxy halos
at sufficiently high redshift $z_i$ that the dark
energy density is negligible. The clustering of these halo centers is
then simply a biased version of the mass distribution. Hence at $z_i$
one can write $\delta_g(z_i) = b(z_i) \delta_M$, with $b(z_i)$
independent of scale for large enough scales. As they evolve
to redshifts below unity, their motions are given by the potential
$\psi$, just as for the matter field. Hence their evolution obeys the
continuity and Euler equations: $\dot{\delta_g} \simeq - \theta_g/a$ and 
$\dot{\theta_g} \simeq -H \theta_g + k^2\psi/a$. The matter density
obeys the same equations with $\delta_M$ and $\theta_M$ as the density
and velocity perturbations. This means that the bias factor preserves
its scale independence: at low redshift, it relates the galaxy power spectrum 
to the matter power spectrum and is not directly sensitive to the clustering of
dark energy. For example the halo model expression \cite{MoWhite,
Cooray2002} for the bias evolution is: $b(z) \simeq
1+(\nu-1)/\delta_{sc}(z)$ where $\delta_{sc}(z) \propto D(z)$
is the density required for spherical collapse at $z$, and $\nu\equiv
\delta_{sc}(z)/\sigma$ with $\sigma$ the smoothed rms mass fluctuation. 
The expression for ellipsoidal collapse has two additional parameters but
still has no scale dependence. 

Clustered dark energy follow Eq. \ref{eqn:DE2}
with $w\neq 0$ and $\sigma\neq 0$, so it has a different time and
spatial dependence from $\delta_m$. 
If the dark energy clusters significantly, 
it is therefore possible that galaxies have a scale dependent bias
relative to it and therefore to the total density field. 

The above argument is very general but relies on some
approximations. These are well justified for massive halos, for
which the evolution at low redshift is very simple: 
consider galaxy halos of mass $M\gg M_*$, where $M_*$ is the standard 
halo model nonlinear mass. The centers of mass of these halos can be
mapped to high-$\sigma$ peaks in the nearly Gaussian mass distribution
at high redshift. Moreover, they do not move significantly, so it is
evident that their power spectrum at large scales evolves simply by
the growth of its amplitude. Such massive halos correspond to
galaxy clusters and LRG's at moderate to high
redshift. For galaxies in lower mass halos, halo motions and mergers
change their  clustering at low redshift, so one has to be careful in
modeling their bias factors. 

Another route to $\delta$ in any GR scenarios is through the metric
potentials. Given lensing measurements of $\psi+\phi$ and dynamical
measurements of $\psi$, one can obtain the potential $\phi$. Using
this, the Poisson equation (\ref{eqn:DE3})  then gives $\delta$, since
the Gravitational constant is known in GR. Thus $\delta$ is 
not independent of the metric potentials even for clustered DE
models. 

\subsubsection{Empirical determination of bias}
To leading order then, knowledge of $b_1$ allows us to relate $P_g$
to $P_\delta$.  Barring extreme scenarios of clustered dark energy,
we take $\delta$ to be the full density field. 

Provided a halo-model description applies reasonably well to our 
universe, bias can be determined by combining
observations and using two and three-point statistics. 
For concreteness we consider the bias parameters $b_1, b_2$ that can be 
determined from the data using the power spectrum and bispectrum
(denoted $B$) measurements. 
In a deterministic bias model, one can then get the density power spectrum. 
With $P_g = b_1^2 P_\delta$ and the reduced three-point parameter $Q
\sim B/P^2$ (see the Appendix and \cite{Bernardeau2002} for full
expressions), one has a relationship between the $Q$ parameter of
galaxies and mass \cite{Frieman1994,Fry1994}:
\begin{equation}
Q_g = \frac{Q_\delta}{b_1} + \frac{b_2}{b_1^2} , 
\end{equation}
By using $P_g$ and measurements of $Q_g$ for different triangles, both bias 
parameters and $P_\delta$ can be determined. 
(A similar 
analysis can be done in real space, e.g. using counts in cells. The skewness 
$S_3$ is given by the shape of the power spectrum and bias parameters.) 
While this is a simplified model, it helps us address what changes for MG: 
the predictions for $P_\delta$ and $Q_\delta$ both change, with the former
given by the new linear growth factor on large scales and the 
latter by next order terms in perturbation theory (see the Appendix for
more details). For well specified
gravity scenarios, these calculations can be done and thus the bias
factors determined from measurements.  

A second approach to measuring $b_1$ is to use the galaxy-mass
cross-correlation measured by galaxy-galaxy lensing in combination
with the galaxy power spectrum (e.g. \cite{Seljak2006}). This has the advantage that one uses only two-point
statistics that can be measured with high accuracy. However, as discussed
below and by \cite{Zhang07}, for MG theories there is a complication
because the Poisson equation is needed as
well since lensing measures the potentials rather than $\delta$. So
for MG theories, the extraction of the bias parameter in this approach
is more complicated -- but nevertheless feasible by jointly fitting
for bias and $\tilde{G}_{\rm eff}$. 

\subsubsection{Galaxy cluster mass function}
A different probe of $\delta$ is provided by
{\it the mass function of galaxy clusters}. Given 
Gaussian initial conditions and a spherical/ellipsoidal collapse model, 
the number density of galaxy clusters can be related to the linear
density contrast.  
In the spherical collapse scenario, a region containing mass $M$ will collapse
if the overall density fluctuation exceeds a threshold $\delta_c$. The number
of such regions can be predicted from the Gaussian statistics and this fixes
the halo mass function $dn/dM$,  the number of halos with mass $M$. 

In the standard $\Lambda$CDM cosmology, gravitational dynamics is
determined by GR. The   
mass function of galaxy clusters is sensitive to the smoothed mass
density variance $\sigma_R^2$ on scale $R$, which is dependent on the
cluster mass and is typically of order 10 Mpc
(e.g. \cite{Pierpaoli2003}). This is related to the density power
spectrum as:  
\begin{equation}
\sigma_R^2 = \int \frac{d^3 k}{(2\pi)^3} \ P_\delta(k) W_{\rm top-hat}^2(kR) , 
\end{equation}
where $W_{\rm top-hat}$ is the window function for averaging with a 
spherical top-hat. 

For clustered DE models, the cluster formation picture becomes
complicated.   The 
presence of the anisotropic stress invalidates the spherical collapse
model and more complicated models such as ellipsoidal collapse 
with tidal fields need to be used. Furthermore, the
fate of an overdense region is no longer determined by the matter fluctuation
$\delta_m$ alone. DE 
fluctuations $\delta_{\rm DE}$ and  $\sigma$ affect $\psi$ through equations
\ref{eqn:DE3} and \ref{eqn:DE4}. And $\delta p$ affects the evolution of
$\delta_{\rm DE}$ through Eqns. \ref{eqn:DE1} and \ref{eqn:DE2}.
Thus a combination of $\delta_m$, $\delta_{\rm DE}$,
$\delta p$ and $\sigma$ act in determining the evolution of a given region of
matter -- the resulting collapse condition has yet to be worked out. 
Since many galaxy clusters form recently at $z\la 1$, where DE is
non-negligible, DE fluctuations could leave some detectable signatures
in cluster abundance. 

For probing the dark universe, this is a valuable feature. It
implies that {\it galaxy cluster 
  abundances contain information on not only fluctuations in
  matter but also fluctuations in dark energy, and thus is a
  promising probe of the total $\delta$}. To extract such information
requires further modeling of the DE model, but a simplified model
can be obtained as follows. The collapse condition based on energy
conservation should be linear in the DM and DE perturbation variables,
since they are all first order 
variables of the energy-momentum tensor. At high redshift, 
the dark energy contribution should vanish (assuming $\bar{\rho}_{\rm
  DE}\ll \bar{\rho}_m$).  
Thus we may assume that  matter fluctuations  are the only source of 
growth for the late time $\delta_m$ as well as $\delta_{\rm DE}$,  $\delta p$
and $\sigma$  responsible for the LSS. In this picture, all perturbation
variables are correlated and have deterministic relations
\cite{Abramo07}. The collapse condition can be simplified into a modified
condition on $\delta_m$ alone.  
An effective  $\delta_c^{\rm eff}$ can be defined for specific DE
models, such that when
a region reaches  $\delta_m\geq \delta_c^{\rm eff}$, it will 
collapse.

The usual collapse model deals with isolated objects and thus 
Birkhoff's theorem is implicitly required. Modifications in GR result
in a generic breakdown of the Birkhoff's theorem. This significantly
complicates  the modeling of cluster abundance in MG models, since the
fate of a given region is determined not only by matter and energy
inside this region, but also matter and energy outside. However,
given a MG model, one can still predict the probability for a given
region with overdensity $\delta_m$ to collapse and thus predict 
cluster abundances.

{\it Assumption/Caveats:} Unlike the use of gravitational lensing to probe
  $\psi+\phi$,  it is model dependent to probe $\delta$ from cluster
  abundance.  (1) The cluster abundance  requires
  careful modeling, even in the simplest case of smooth DE models. For
  example, the tidal field makes the spherical collapse model only 
  a rough approximation. (2) The observable-mass relation is needed 
  to connect observable (e.g. X-ray flux,
  SZ flux or cluster richness) to the mass of clusters. These cluster
  properties often involve complicated gastrophysical processes and
  can not be predicted with sufficiently high precision from first
  principles. As a consequence, using cluster abundance to probe
  $\delta$ often require model-dependent calibrations.    

In spite of these caveats, it may be hoped that the well posed problem
of the evolution
of a region in an initially Gaussian random field 
will be calculable, and related to the linear density field in
generic MG or DE models. 

\bigskip
{\bf Velocity divergence $\theta$:} 
Many existing velocity measurement are based on distance
indicators: the difference of the true distance from what is inferred from 
the recession velocity gives an estimate of the peculiar velocity of a sample 
of galaxies or clusters \cite{FP}.   The pairwise velocity at small
separation can be measured through anisotropic galaxy clustering in
redshift space at  cosmological distances \cite{v12}. 
While challenging, there are ongoing attempts to improve measurements of 
 bulk flow measurements, based on SNe Ia
  \cite{SNIameasurement}.  An independent method 
is the kinetic Sunyaev Zel'dovich
(SZ) effect \cite{Sunyaev80} of clusters which is directly proportional to the
cluster peculiar velocity and enables a rather model independent measurement
method \cite{KSZ}.  These  measurement are likely to have lower
signal-to-noise  than the redshift space distortions discussed above. 
Further it is unclear whether they estimate $\theta$ of the total
fluid in a clustered DE
scenario, for the reason discussed above. 

\bigskip
{\bf CMB: } 
The CMB power spectrum is given by: 
\begin{equation}
C_{TT}(l) = \int dk \int d\chi'\ F_{CMB}(k, l, \chi')\ j_l[k\chi(z')]
\end{equation}
where the spherical Bessel function $j_l$ is the geometric term
through which the CMB power spectrum depends on the distance to the last 
scattering surface. The function $F_{CMB}$ 
combines several terms describing the primordial power
spectrum and the growth of the potential. We will regard $F_{CMB}$ as
identical to the GR prediction since we do not invoke MG in the early
universe (up to the redshift at last scattering). 

The CMB anisotropy 
does receive  contributions at redshifts below last scattering, in
particular due to the 
integrated Sachs Wolfe (ISW) effect \cite{ISW}. In the presence of dark
energy or due to modifications in gravity, gravitational potentials
are in general time varying and thus produce a net change in the
energy of CMB photons: 
\be
\frac{\Delta T}{T}\left.\right|_{\rm ISW}=-\int
\frac{d(\psi+\phi)}{dt} a(t) d\chi \ \ .
\ee
The ISW effect, like gravitational lensing, depends on and probes  the
combination $\psi+\phi$. The ISW signal is overwhelmed by the primary CMB at
all scales (although it does produce a bump at the largest scales in 
the CMB power
spectrum). For this reason, it has to be measured indirectly, through
cross-correlation with other tracers of large scale structure. The
resulting cross-correlation signal is then
\be
C_{\rm ISW}(l)=\int P_{g
  (\dot\psi+\dot\phi)}(k=\frac{l}{\chi},\chi)\ a^2\ 
\frac{d\chi}{\chi^2}\ .
\ee
Here, $P_{g(\dot\psi+\dot\phi)}\left( k,\chi\right)$ is the
cross-power spectrum of $(\dot\psi+\dot\phi)$ and galaxies or other
tracers of the LSS such as quasars or clusters.  By cross-correlating
the CMB temperature with galaxy over-density $\delta_g$, the 
ISW effect has been detected at $\la 5\sigma$ confidence level
\cite{ISWmeasurements}  and provides independent evidence
for dark energy, given the prior of a spatially 
flat universe and GR. This cross
correlation signal depends on galaxy bias, which has to be marginalized to
infer cosmology. With the aid of gravitational lensing, uncertainties of
galaxy bias can be avoided \cite{Seljak99,Zhang06b}.
Furthermore, since the ISW 
amplitude peaks on the largest scales, it also has a strong correlation with
large scale bulk flows and produces a cross correlation signal with
potentially better signal-to-noise 
than that of the density-ISW cross correlation \cite{Fosalba07}. 

The primary CMB is Gaussian and statistically isotropic. However, gravitational
lensing distorts  the CMB sky and induces anisotropy and 
Fourier mode-coupling in the CMB, which should not exist otherwise. This
feature should allows reconstruction of the lensing potential from future high
resolution CMB maps \cite{CMBlensing}.  The CMB sky is the furthest
lensing source and thus can probe $\psi+\phi$ at redshifts well above
unity. This will be useful to constrain those MG and DE models in
which deviations from $\Lambda$CDM persist at these redshifts.

ISW measurements and  future measurements of lensing and
galaxy clustering can probe scales approaching the horizon scale. This 
provides an additional test of MG models in which the growth of
perturbations is altered at relatively high redshift on super-horizon
scales. \cite{Bertschinger2006} 
showed that growth on super-horizon scales is constrained to be
universal for MG models with $\psi=\phi$. \cite{Hu07} show how it
differs for $f(R)$ models which do not obey this constraint, and 
describe the transition from super-horizon to sub-horizon scales. 
If measurements achieve high accuracy on these large scales, they can
be combined with information on sub-horizon scales to provide
additional constraints on the ratio of potentials $\eta$ for such MG
models. 

{\bf Summary:} The quantity that can be measured most
robustly is the sum of potentials $\psi+\phi$, through
gravitational lensing and the ISW effect. With a bit more modeling,
the Newtonian potential $\psi$ can be inferred from galaxy
velocity measurements (i.e. redshift space distortions). 
To obtain model independent constraints on the total density
perturbation $\delta$ is challenging if one allows for dark energy
clustering in the GR scenario. Galaxy
clustering is likely to be an effective 
measure of the matter fluctuation $\delta_m$, while cluster abundance is
a promising probe of $\delta$ as it is sensitive to DE fluctuations as
well. Although the galaxy peculiar velocity is likely to be well
measured in the future,  the DE peculiar velocity (and therefore the
overall $v$ and $\theta$) is likely 
the most difficult to measure. Cross-correlations of large-scale
structure tracers with the lensing potential or the Newtonian
potential are probably the most promising tests of MG in the near
future, as we discuss next.

\subsection{Joint constraints from multiple observations}
\label{sec:joint}

If multiple observables are to be combined, model
independent information can only be inferred if they probe the same
range of redshift and length scale. 
The distance-redshift relation will be
measured to $\sim 1\%$ accuracy  by the next generation SNIa and BAO
surveys at low-z and by the CMB at high-z. The next generation BAO
surveys can further measure $H(z)$ at low redshift. 
With the expansion rate of DE and MG models tightly
constrained, measurements of perturbed variables become powerful
discriminators.  

The distance-redshift relation at redshift $z$ is given by an
integral over the expansion rate, and therefore the energy densities,
from redshift 0 to $z$. This measurement at $z\simlt 1$ has provided
evidence of acceleration, consistent with $\Lambda$CDM. On the other
hand, CMB measurements at high-$z$ for both distances and perturbations
are consistent with a universe governed by GR, with its energy density
dominated by matter and radiation \cite{Hu2004}. 
Thus either dark energy or modification of gravity must produce effects 
that are significant at $z\simlt 1$ and negligible at $z\sim 1000$. 
In Fig. \ref{fig:DistanceGrowth} we show as examples the deviation 
(from $\Lambda$CDM with $\Omega_{de}=0.7$) of a model with 
$\Omega_{de}=0.75$ and a flat DGP model with $\Omega_m=0.3$.  It is
clear that for both 
distances and perturbations, significant deviations occur at low-z in
such models 
\footnote{There do exist both DE and MG models with observational
consequences at redshifts significantly higher than 1: these include 
oscillating $w$ models and TeVeS (see also \cite{Hu07,Acquaviva2007}). 
Future surveys will provide some 
probes of this higher redshift universe through effects such as
CMB lensing, high-z galaxy surveys and 21 cm redshift space
measurements. } . 

The most promising scale/redshift range in the near future is 
$\sim 10-100$s Mpc at redshifts $\sim 0.3-1$. Imaging and
spectroscopic observations are likely to be made on these scales and
will be robust to many sources of error and dependence on specific
models.  We list below several categories of 
surveys that will test MG and DE models. Two sets of surveys are
indicated: surveys planned for the near future (significant data 
within 5 years), and surveys planned to start in about a decade. (The
list is not complete as several projects have been formulated
or modified recently.)

\bfi{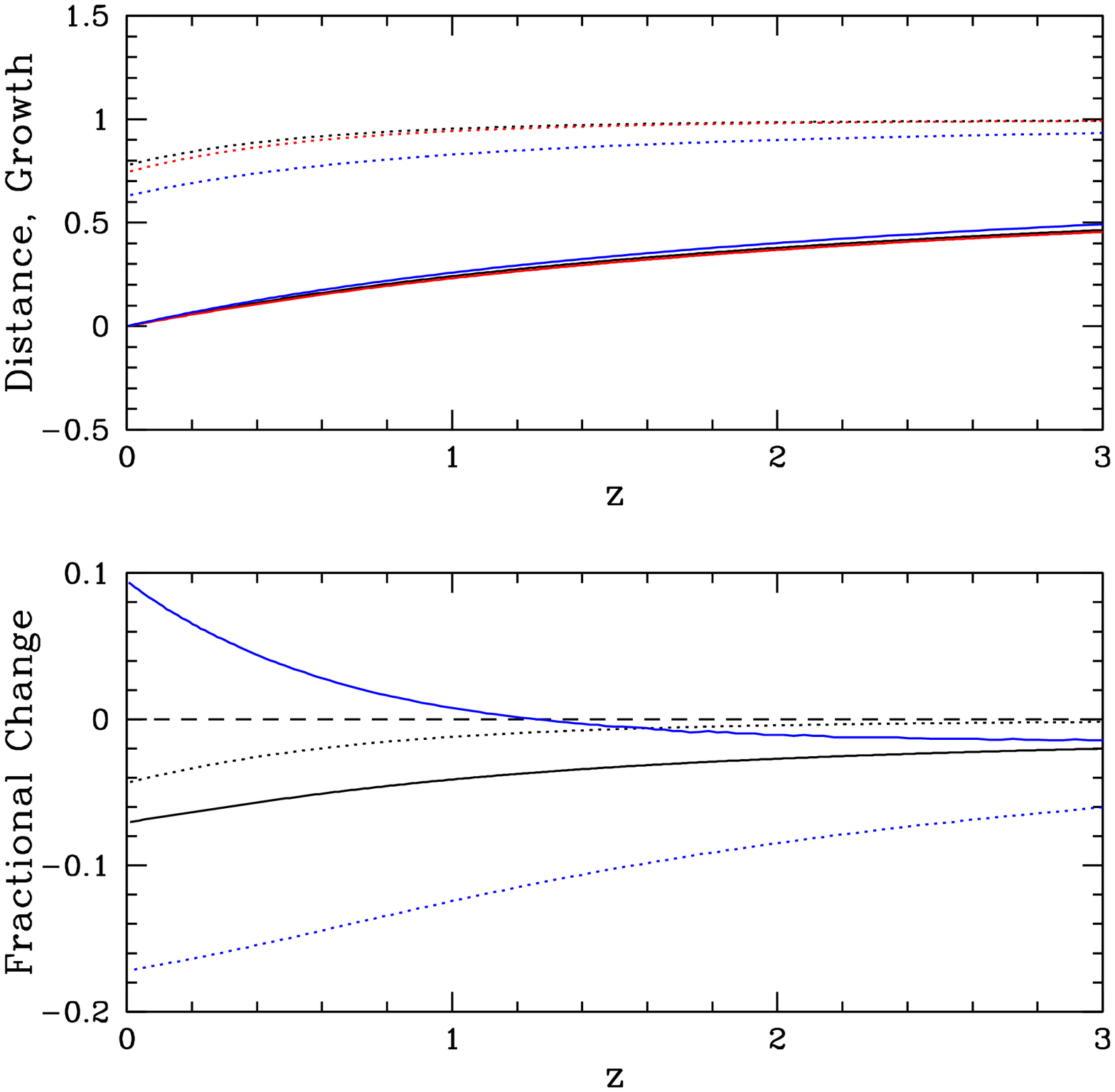}
\caption{
{\it Upper panel:} 
We plot the normalized distance $d(z)$ (solid curves, almost coincident) 
and the linear growth rate $D(z)/a(z)$ (dotted curves) 
for $0<z<3$ for two dark energy models (in black and red)
and a DGP model (in blue). The distances and growth rates are 
normalized to give 1 at high redshift (z=1100). 
{\it Lower panel:} 
The fractional deviations in distance (solid curves) and linear growth
(dotted curves) from the fiducial $\Lambda$CDM model are shown for 
a dark energy model (black) and DGP (blue) (see text for details). 
Note that the DGP growth curve is 
sensitive to the parameters chosen to fit the distance redshift
relation: for a distance curve that matches $\Lambda$CDM better, the growth
curve would also have smaller deviation. 
\label{fig:DistanceGrowth}}
\efi

\begin{itemize}
\item Multicolor imaging survey: With photometric redshifts for
  millions of galaxies, these surveys provide measurements of weak 
  lensing, galaxy cluster abundances, and the angular clustering of
  galaxies, clusters and quasars. These measurements probe
  $\psi+\phi$, $\delta_g$ and their cross-correlation. Upcoming
  surveys include: 
\begin{itemize}
\item DES, KIDS, PS1, HSC (2008-). $0<z\simlt 1$.
\item LSST, SNAP, DUNE (2014-). $0<z\simlt 3$. 
\end{itemize}

\item Spectroscopic surveys: While primarily designed to measure
the distance-redshift relation and $H(z)$ using the baryon acoustic
oscillations in galaxy power spectra, they will provide improved 
measurements of $P_{g}$, $P_{gv}$, $P_{vv}$ on large scales. Some
surveys will target $z\simlt 1$ galaxies and others will select
galaxies at higher redshift, $2\simlt z\simlt 3$. 
\begin{itemize}
\item LAMOST, 
WiggleZ, HETDEX, WFMOS, BOSS (2008-). $0<z\simlt 3$. 
\item ADEPT (2014-). $1\simlt z\simlt 2$. 
\end{itemize}

\item 21 cm surveys: 
SKA \footnote{http://www.skatelescope.org/}(2015-). The square
kilometer array (SKA) has the potential to 
detect $\sim 1$ billion galaxies over $0<z\simlt 1.5$, with a deeper
survey extending to $z\sim 5$, through 21cm line
emission of neutral hydrogen in galaxies. If successful, it will 
provide high precision measurements of the distance-redshift relation
through BAO's \cite{Abdalla05}, and tests of MG through: 
(a) weak lensing maps 
with accuracy comparable to that of large optical surveys 
\cite{SKAlensing}, (b) velocity measurements 
through redshift distortions of galaxy clustering \cite{Zhang07}, 
and (c) ISW measurements through CMB-galaxy cross-correlations. 

\item SZ and X-Ray Cluster surveys: These will measure the abundances
  of galaxy clusters out to $z\sim 1$ and beyond. Cosmological
  applications will depend on supplementary optical data to get
  photometric redshifts of the detected clusters. 
\begin{itemize}
\item SZA, SPT, ACT, APEX, eROSITA (2008-). $0< z \simlt 1$. 
\end{itemize}

\item CMB: temperature and polarization maps provide high-z
  constraints and also measurements of the ISW effect and CMB lensing,
  which are probes of $\psi+\phi$ at lower redshift. 
\begin{itemize}
\item PLANCK and ground based missions (2008-)
\end{itemize}

\end{itemize}

We have indicated the approximate redshift range over
which these surveys will provide accurate measurements. 
It would be most useful to have different observables overlap in
redshift and length scale in the range 
$z\simeq 0.3-1$ and at scale $\lambda\simeq$ 10 to several 100
Mpc. This range of scales covers the linear and quasilinear regimes of
structure formation (we are assuming that MG effects are present on
these scales). We consider next two promising combinations of observables
that on the 5-7 year timescale will enable measurements of the MG
functions $\tilde{G}_{\rm eff}$ and $\eta$ on these scales. 

{\bf Lensing and galaxy power spectra:} 
Planned next-generation imaging surveys (see above)
will have area coverage in excess of $1000$ sq. degrees, enabling
few percent level measurements of lensing power spectra. 
The same imaging surveys will also measure the
angular clustering of the galaxies $C_{g}$ (at $z\sim 0.3-0.6$, 
the redshift of the lensing mass) to percent level accuracy; cluster
abundances will also be 
measured through optical and SZ surveys: both measurements probe
the matter density $\delta$. Alternatively, spectroscopic surveys like
BOSS will measure $P_{g}$, the three-dimensional power spectrum, to
percent level accuracy. The shear power 
spectra can be combined with the density power spectra measured 
at $z\sim 0.3-0.6$ and scales of 10-100 Mpc. Using the Poisson equation, 
$\tilde{G}_{\rm eff}$ will then be tightly constrained, assuming statistical 
errors dominate the error budget. The main galaxy sample and LRG
sample of the SDSS has already been used in constraining MG models
through galaxy clustering alone (e.g. \cite{Shirata2007}), though in a
model-dependent way.  

{\bf Cross-correlations of galaxies with shear and velocity:}
Galaxy-galaxy lensing measurements made from imaging surveys 
probe the lensing potential-galaxy correlation.  This measurement has
been made to high accuracy from the SDSS \cite{Sheldon2004,Mandelbaum2006}. 
In the near future one can expect measurements of 
$P_{gv}$ to a few percent (e.g. from the BOSS survey) at $z\sim
0.3-0.6$. In combination with percent-level galaxy-galaxy lensing
measurements over the same range of redshift,  
the ratio of potentials $\eta$ will be precisely constrained \cite{Zhang07b}. 

The measurements described above would be major advances in
constraining MG theories, as the current constraints on $\sim 10-100$ Mpc
scales are weak and insufficient to test MG theories with
any robustness. For particular models the
scale and redshift evolution of a single statistic, such as the
lensing power spectrum, can be powerful as well. 
We leave for future work a detailed study of how well
these measurements will test MG theories. 
In considering an observable suitable for distinguishing
models of gravity, one must address the familiar problems in extracting 
cosmological information due to statistical and systematic 
errors, i.e. the expected precision on the measurement, 
the physical assumptions necessary to
connect observable to the four variables of interest, and the degeneracy
with other cosmological parameters. 

\section{Modified gravity vs. dark energy}
\label{sec:DGP}

Specific models of MG and DE can be tested by 
combining observations of the expansion rate and large scale
structure. For example, in the $\Lambda$CDM model (and well defined scalar
field models), the growth of the large scale structure is completely
determined by the expansion history: there exists a fixed relation
between the expansion rate and the growth of LSS. This consistency 
check has been carried out in the
literature and is indeed able to distinguish specific models investigated
\cite{consistencycheck}. Furthermore, this consistency check can be
performed in a  model independent way to search for signatures of
violation of GR, with the prior of smooth DE \cite{Zhang03}. Current data
pass this consistency check, although violations of the consistency
relation can not be ruled out \cite{Wang07}.

DE models that depart from scalar field models can be much more  complicated,
with a break down of the correspondence between the expansion rate 
and large scale structure.   The expansion rate is
determined by $\bar{\rho}_{\rm  GR}$ and $w$. However, two extra DE
properties, the anisotropic stress $\sigma$ and the response of
pressure to perturbations ($\delta p$), can affect the growth of the
LSS.  These two properties are determined by the microphysics of
the DE model and  are independent of $\bar{\rho}_{\rm  GR}$ and $w$. As
a consequence, the growth of LSS is no longer fixed by the expansion rate
and  the above consistency check can not be applied to search for
signatures of the violation of GR. Current observational constraints on
the sound speed ($c_s^2\equiv \delta p/\delta \rho$)
\cite{soundspeed} and the anisotropic stress \cite{anisotropicstress}
are weak. Furthermore, these studies use a particular form of
$\sigma$ and $\delta p$ and assume that one can be switched off
when studying the other.  Thus a potentially wide range of DE models
with non-negligible anisotropic stress and pressure fluctuations are
still viable against observations. To investigate the feasibility of
distinguishing between DE and MG, we will allow for arbitrary
anisotropic stress and pressure perturbations. 

Modifications of gravity (at least the class of
theories we have considered) involve two extra
quantities which govern LSS, namely, modification of 
Newton's constant, ${\tilde{G}_{\rm eff}}$, and the ratio of
potentials $\eta\equiv \phi/\psi$. Although these quantities
determine the gravitational interaction of perturbations, they do 
in general affect the expansion rate $H(z)$ -- unlike for GR and 
its Newtonian limit. This is in part due to the fact that 
for MG Birkhoff's theorem no longer holds and thus
the usual exercise of calculating $H(z)$ from the Newtonian dynamics of
a spherical matter distribution no
longer applies. Thus ${\tilde{G}_{\rm eff}}$ and $\eta$ 
represent real extra degrees of freedom in MG theories. 

The extra degrees of freedom in MG and clustered DE models  can produce similar
observational consequences. For 
example, the anisotropic stress breaks the equality between $\phi$ and
$\psi$, mimicking the role of $\eta$ in MG models. Thus
one might expect that by tuning the two 
extra degrees of freedom in DE models, one can  mimic a
given MG model  to fit observations.  Indeed,
Kunz \& Sapone  \cite{Kunz06} explicitly construct a DE model
which reproduces degenerate $\phi$, $\psi$ and $\delta_m$
with the flat DGP modified gravity model.   
\footnote{Notice that this DE model has large dark energy fluctuations
on scales well below the horizon,  so it differs from 
conventional clustered DE models that rely on scalar field dynamics. 
Furthermore, usually $\delta p$ is
parameterized as $\delta p=c_s^2\delta \rho$ ($c_s$ can be both scale
and time dependent). The DE model considered by \cite{Kunz06} has an unusual
  form of pressure perturbation, in which $\delta p$ is connected to
  the anisotropic stress $\sigma$ instead of $\delta$.  } 

This degeneracy certainly deserves further investigations. In
this section, we consider in more details the question: can one
always succeed in  tuning DE models to produce observational
consequences identical to a given MG model?  {\it If the answer is yes, then
one can never unambiguously test for deviations from GR}. 

The answer to the above question is incomplete in fully describing the
dark degeneracy. The complementary question, which needs to be answered
is: can one always tune MG models to  produce observational
consequence identical to a given DE model?  {\it If the answer is yes, then
  one can never unambiguously justify the existence of DE}. However,
this question is more difficult 
since it requires a general parameterization of the relation between
the expansion rate  $H(z)$ and  the nature of a general gravity
theory -- such a parameterization is not yet available. This limit in
theoretical understanding of MG forces us to investigate only the
first question, since we know the most general way of parameterizing
the influence of DE on the expansion history of the universe.
Furthermore, we have constrained our study to a special class of MG 
models,  in which gravity is minimally coupled to matter.  The study of
both questions for the most general MG models is beyond the
scope of this paper.

The relationship between the four perturbation variables $\phi$,
$\psi$, $\delta$ and $\theta$ is fixed for
a complete DE or MG theory. These consistency relations are the key to
probing the nature of 
DE and MG. With just two variables being observable, one can only test
against one consistency relation and, as we see below, 
by tuning the two extra degrees of freedom in clustered DE models, any MG
model can be 
mimicked. However, with more observed variables, one can test other
consistency 
relations and hope to break the degeneracy between DE and MG
models. In this section, we explore the feasibility of distinguishing DE
and MG models. 
In  this section we consider only the question of distinguishability in
principle, without regard to the accuracy of observations in the
foreseeable future. 

\subsection{Two perturbation observables}
\label{sec:twovariables}
{\it First we assume that both potentials are observables,
i.e. we require $\phi$ and $\psi$ to be identical in the two
models}. From the discussions in previous sections, these two quantities
are the most likely to be measured to high precision. So we set them
identical in the constraint equations for 
GR and MG to get relations between the remaining variables. 
Comparing Eqn. \ref{eqn:DE4} for $\sigma$ in GR with the
constraint Eqns. \ref{eqn:MG3} and \ref{eqn:MG4} for MG gives
\be
\label{eqn:constraints1}
\sigma=\frac{2}{3}\frac{\eta^{-1}-1}{\eta^{-1}+1}\frac{\tilde{G}_{\rm eff}}{G}\frac{\bar{\rho}_{\rm
 MG}}{\bar{\rho}_{\rm GR}} \delta_{\rm MG} \ .
\ee

In addition by combining the Poisson equation (\ref{eqn:DE3}) for GR with 
Eqns. \ref{eqn:MG3} and \ref{eqn:MG4} for MG, we obtain  a second
constraint 
\be
\label{eqn:DE5}
\delta_{\rm GR}+3(1+w)Ha\frac{\theta_{\rm GR}}{k^2}=\frac{2}{\eta^{-1}+1}\frac{\tilde{G}_{\rm
    eff}}{G}\frac{\bar{\rho}_{\rm 
 MG}}{\bar{\rho}_{\rm GR}} \delta_{\rm MG}\ .
\ee 

The question then is whether Eqns. \ref{eqn:DE1},  \ref{eqn:DE2}
and \ref{eqn:DE5} have solutions for $\delta_{\rm GR}$, $\theta_{\rm
  GR}$ and $\delta p$, in terms of MG variables (recall that
$\sigma$ is now fixed by Eqn. \ref{eqn:constraints1}).  Without a
fundamental theory,  $\delta p$  can take any form \cite{Hu99}: hence
there is {\it always} a form of $\delta p$ satisfying all three
equations. Namely, there is always a DE model which can mimic the
given MG model to produce identical $\phi$ and $\psi$. 

The degeneracy persists for other combinations of two
perturbation variables. We have discussed above that in a clustered DE
model, it is difficult 
  to establish whether galaxies, other tracers or cluster abundances
  probe $\delta$ or 
  $\delta_m$ (or neither!). If we assume that a sub-set of LSS
  observations will provide measurements of $\delta$, then combining
that with measurements of  $\psi+\phi$ from lensing, we have
\begin{equation}
\label{eqn:constraints2}
\sigma=\frac{2}{3(1+w)}\delta\left(\frac{\tilde{G}_{\rm eff}\
  \bar{\rho}_{MG}}{G \ \bar{\rho_{\rm GR}}}-1\right) \ .
\end{equation}
Thus if $\sigma$ is free, it can be chosen to match the above equation
for any set of theories. 

Extra information can  break this degeneracy. The response of pressure
to density perturbations and the anisotropic stress are determined by the
microphysics of the DE model. It requires a theory to provide
such closure relations (see \cite{Hu07} for more detailed discussions). 
For example, the quintessence 
model predicts vanishing $\sigma$ and negligible pressure perturbation
on sub-horizon scale. Even if advances in the understanding of general DE
theory do not provide such specific information, some general
constraints can still break the degeneracy. 
For example, if $\delta p$ takes the form
$\delta p=c_s^2 \delta \rho $ and
$c_s^2=c_s^2(t)$, as is true for the adiabatic case, solutions do not
 exist in general for equations 
 \ref{eqn:DE1},  \ref{eqn:DE2}  and \ref{eqn:DE5}. In this case, one
can not find a DE model to mimic the given MG model. 

Another physically well motived example is for the anisotropic stress
$\sigma$. A natural source of  $\sigma$ is the velocity perturbations in the
fluid.  By the requirement on  gauge invariance, the evolution in $\sigma$
may be parameterized in the following form in the Newtonian gauge
\cite{Hu98,anisotropicstress},
\be
\label{eqn:neutrino}
\sigma+3H\dot{\sigma}=\frac{8}{3}\frac{c^2_{\rm vis}}{1+w}\theta \ ,
\ee
where $c_{\rm vis}$ is the viscous parameter.
This equation in general contradicts equation
\ref{eqn:constraints1} and \ref{eqn:constraints2} above and thus no DE model
that satisfied Eq. \ref{eqn:neutrino} can mimic the given MG model. 

Extra information can also come from additional observables. The 
equations above show that if we have
just one additional observable, such as $\delta$ or $\theta$,  there will in
general be no solution for the remaining two variables that satisfies
three equations (e.g. \ref{eqn:DE1},  \ref{eqn:DE2}  and \ref{eqn:DE5}).  We
consider this next.

\subsection{Three or more observables}
\label{sec:threevariables}
\bfi{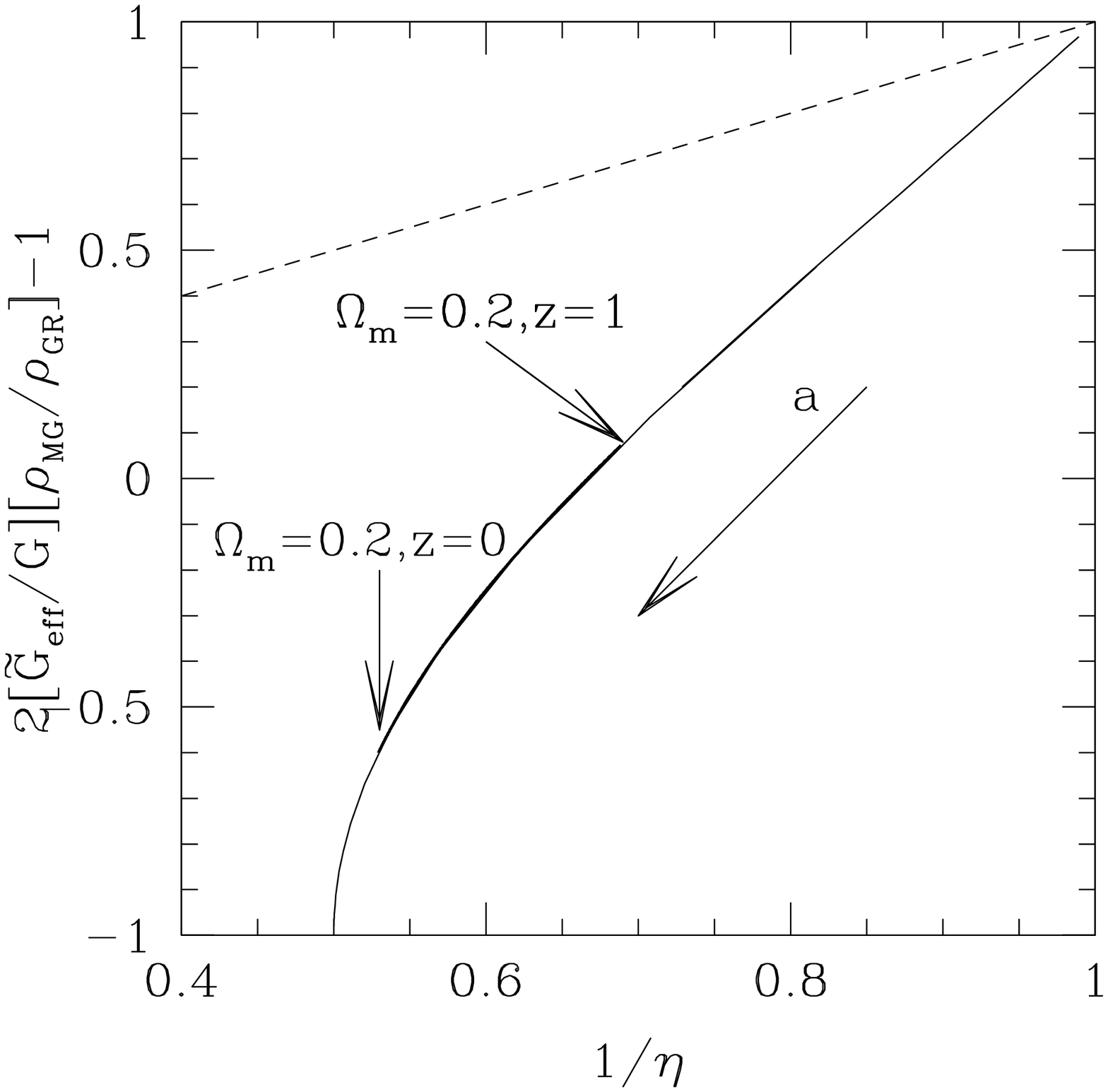}
\caption{First consistency condition for at least one  DE model to 
  mimic $\phi$, $\psi$ and $\delta$ in a flat DGP
  model. The dashed line given by Eqn. \ref{eqn:eta_Geff}
  represents the required
  condition, while the solid curve is the actual  relation in flat
  DGP. When $a\rightarrow 0$, $\eta\rightarrow 1$ and for $a\rightarrow
  \infty$, $\eta\rightarrow 1/2$. The points on the curve with $a=0.5$
  and $a=1$ are indicated. (For flat DGP
lines with different $\Omega_m$ lie on top of each
other.) The disagreement between the two curves shows that  DGP is a
modified gravity  model that can not be mimicked by any dark energy model. 
\label{fig:MG}}
\efi

If both potentials and $\delta$ are observable then the theory is
constrained much more tightly, especially if they are measured multiple
redshifts.  

\bfi{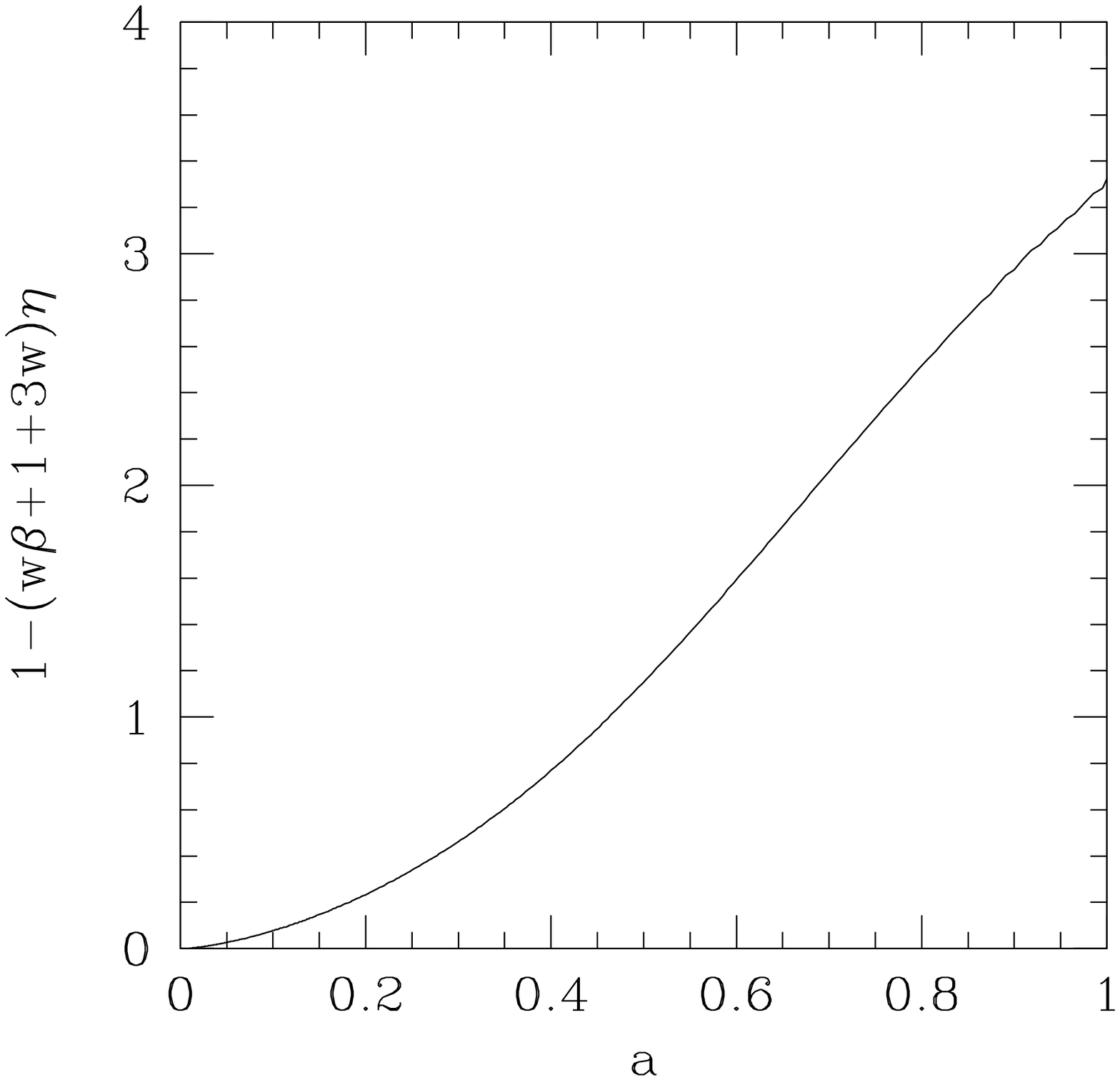}
\caption{The second necessary condition for at
  least one  DE model to mimic $\phi$, $\psi$ and $\delta$ in the given DGP
  model. The condition $\eta^{-1}\simeq w\beta+1+3w$, given by 
  Eqn. \ref{eqn:eta} implies the variable on the y-axis should be zero
  for all $a$. For flat DGP with $\Omega_m=0.2$, this
  condition is also severely violated for $a>0$. \label{fig:MG2}}
\efi

For a DE model to mimic the given MG model, 
 $\delta$, $\phi$ and $\psi$ must satisfy the three equations
 \ref{eqn:DE3}, \ref{eqn:MG3} and  \ref{eqn:MG4}. This imposes the 
the following consistency relation for
$\tilde{G}_{\rm eff}$ and $\eta$, 
\be
\label{eqn:eta_Geff}
\eta^{-1}=2\frac{\tilde{G}_{\rm eff}}{G}\frac{\bar{\rho}_{\rm
 MG}}{\bar{\rho}_{\rm GR}}-1\ .
\ee
So if the given MG model does not obey the above relation,
no DE model can produce $\delta$, $\phi$ and $\psi$ identical to the
given MG model, no matter how the DE properties are fine-tuned. 

Eq. \ref{eqn:eta_Geff} represents a strong constraint on MG models as
it shrinks the 2-parameter $\eta$-$\tilde{G}_{\rm eff}$ space in MG
models into a straight line.   We show as an example that the DGP
model does not satisfy this condition.  

In a flat DGP model, 
 $\tilde{G}_{\rm eff}=G$ and $\eta^{-1}=(1+1/3\beta_{\rm
 DGP})/(1-1/3\beta_{\rm  DGP})$ \cite{Koyama06}. Here $\beta_{\rm
 DGP}=1-2r_cH(1+\dot{H}/3H^2)=1-2r_cH/(1-2r_cH)<0$, with 
 $H^2=H/r_c+\Omega_m a^{-3}$ and  $r_c=1/(1-\Omega_m)$. We have
 normalized $H(z=0)=1$. For the DGP 
 model, $\bar{\rho}_{\rm MG}=\Omega_ma^{-3}$ (up to a normalization,
 which is irrelevant for this discussion). By requiring the
 DE model to reproduce the expansion history of the given MG model, we
 have $\bar{\rho}_{\rm GR}=H^2$. Fig. \ref{fig:MG} shows that the
consistency condition is significantly violated by all flat DGP models. This
 means that no DE model can produce 
$\phi$, $\psi$ and $\delta$ identical to a flat DGP model that
satisfies observational constraints on the expansion history.

This conclusion seems to contradict \cite{Kunz06}. However,
\cite{Kunz06} require
the dark matter density fluctuation $\delta_m$ 
to be identical in the GR and MG scenarios. We
require the total $\delta$ to be identical instead. In GR, the Poisson
equation specifies the $\phi$-$\delta$ relation, not the
$\phi$-$\delta_m$ relation, so with $\delta$ as an observable, one 
directly construct consistency relations and thus distinguish between DE
and MG.

There is also a constraint on the anisotropic stress, from
Eq. \ref{eqn:DE4}, \ref{eqn:DE5}, \ref{eqn:MG3} and \ref{eqn:MG4}, 
\be
\sigma=\frac{\eta^{-1}-1}{3(1+w)}\delta\ .
\ee
Comparing Eq. \ref{eqn:DE2} with Eq. \ref{eqn:MG2}, we obtain
\be
\theta\left(3Hw-\frac{\dot{w}}{1+w}\right)+\left(\frac{c_s^2}{1+w}\delta-\sigma\right)\frac{k^2}{a}=0
\ .
\ee
Since $H\theta\sim \beta aH^2\delta\ll k^2\delta/a$, we have
\be
\label{eqn:cs1}
c_s^2\simeq \frac{(1+w)\sigma}{\delta}=\frac{\eta^{-1}-1}{3}\ .
\ee
Comparing Eq. \ref{eqn:DE1} with Eq. \ref{eqn:MG1}, we obtain
\be
\label{eqn:cs2}
c_s^2=\frac{w(\theta/a-3\dot{\phi})}{-3H\delta}+w\simeq
w(\frac{\beta}{3}+1)\ .
\ee
Combining both constraints on $c_s^2$, we obtain
\be
\label{eqn:eta}
\eta^{-1}\simeq w\beta+1+3w \ . 
\ee
$w$ is fixed by the condition $\bar{\rho}_{\rm GR}=H^2$. $\beta$ is
calculated from the given MG theory. So the above equation can be
checked from the viewpoint of MG models unambiguously. 

Again, Fig. \ref{fig:MG2} shows that this condition is severely violated for
the DGP model: thus no DE model can mimic a flat DGP model to reproduce
identical $\phi$, $\psi$ and $\delta$.  We have verified that this
is true for $f(R)$ models as well. 

These relations present general constraints, without resort to real
observation data. Observations show that at the present epoch,
$w<-1/3$ since the universe is accelerating, while $\beta>0$ 
since structure is growing. From Eq. \ref{eqn:cs2}, we have
$c_s^2<-1/3$, if the related DE model can reproduce $\phi$, $\psi$ and
$\delta$. Furthermore, Eq. \ref{eqn:eta} tells that $\eta<0$ today.

\section{Discussion}

We have described the role of perturbations in testing theories 
of modified gravity (MG) against large-scale structure observations.  
We have chosen the class of MG theories that are described by scalar
perturbations, so that two metric potentials suffice to describe the
perturbed space-time. We then consider the quasi-static, Newtonian limit
of the perturbed  field equations and compare dark energy (DE) and MG
theories. 

Our main focus is on the relationship of different observables -- lensing,
large-scale dynamics of galaxies, galaxy clustering, cluster
abundances and various cross-correlations -- to the four perturbation
variables of MG theories:  
the two metric potentials, the density field, and the divergence of
the peculiar velocity.  In \S \ref{sec:LSS} we give the relationship
of measured power spectra in real space, redshift space and on the sky with
theoretical predictions for these perturbation variables.  
We highlight the use of
two effective functions to test MG theories: the ratio of the metric
potentials $\phi/\psi$ and the effective Gravitational constant 
$\tilde{G}_{\rm eff}$. We also consider in the Appendix quasilinear
signatures of MG and show how the MG functions affect second order
corrections to the power spectrum and the bispectrum. 

We discuss in detail what is actually measured by various large-scale structure
observations once the assumptions of smooth dark energy and GR are
dropped. While lensing and dynamical probes have a direct connection
to different potential variables, tracers of the density field and
cluster abundances must be treated carefully in extracting information
about the density field from them. The most robust tests of 
MG effects can be made by combining different
observables from planned multi-color imaging surveys and redshift
surveys: see \S \ref{sec:joint} for two examples that will be feasible in the
near future.  

Observables that may not be useful in constraining smooth DE
can be crucial for testing MG because there are more
variables to be measured and different observables are sensitive to
them. For instance, the redshift space power spectrum $P_{g\theta}$
is not considered a valuable probe of structure formation in DE studies
as other methods produce lower statistical errors on dark energy
parameters.  But in a MG 
scenario, $P_{g\theta}$ is useful because it probes the Newtonian potential
$\psi$ that other probes are not sensitive to. It can be combined with
galaxy-galaxy lensing, which probes $P_{g(\psi+\phi)}$, to constrain
the ratio of potentials $\phi/\psi$ (\S \ref{sec:joint}). 
This is a case where observables from
multi-color imaging surveys and spectroscopic surveys must be combined 
to test MG theories. More generally, once one allows for MG scenarios, 
multiple observables are needed to test theories. 
Thus the diversity of LSS observations, which has lost some of its
appeal in the recent trend of going after a single dark 
energy figure of merit, becomes vital (see \cite{White2007} for a
broader criticism of dark energy driven research, and \cite{Kolb2007}
for a rebuttal).  

Finally we consider a question posed recently in the literature: can a
DE model be constructed to mimic any MG theory? We show that with
observations of multiple perturbed variables (three are sufficient in
general), unique signatures of MG theories can be established. We show
with the example of the flat DGP model how, given sufficiently
accurate measurements of the lensing and Newtonian potentials and the
density, no DE model can mimic the DGP model.  

Our results may be compared with other recent studies in the
literature, some of which appeared while this work was in progress
\cite{Kunz06,Hu07,Bashinsky2007}. These studies tackle the question of
how dark energy and modified gravity can be distinguished. We have 
clarified the apparent conflicts between this paper and \cite{Kunz06} in \S
\ref{sec:threevariables}. \cite{Bashinsky2007,Hu07,Pogosian2007} present a 
formal argument that any deviation from GR can be absorbed into the
dark sector as an effective dark component.  The effective stress-energy
tensor of this component, $T_{\mu\nu}^{\rm eff}$, is
defined as the deviation of  the given  MG theory from Einstein's
field equations. $T_{\mu\nu}^{\rm eff}$ is conserved, as expected for
the usual DE model.  From this argument, one might 
conclude that MG can not be distinguished from DE gravitationally. However,
\cite{Hu07} also pointed out that the effective dark component has a
generic, though implicit,  coupling to matter, despite the
conservation of $T_{\mu\nu}^{\rm eff}$.  This coupling hides in the
closure relations for this dark component, which  depend on
external matter and the metric, instead of just its internal 
microphysics. A theory with such a dark component is 
fundamentally different from conventional DE models of the kind we have
considered, even ones with strong clustering of the DE. (Indeed, it
should not be surprising that allowing for a dark component with an arbitrary
stress-energy tensor {\it and} couplings to matter can mimic any
modification of gravity.) 
Our result, that DE models -- with no coupling to matter -- 
can be distinguished from MG, appears to be consistent with the
analysis of \cite{Hu07}. In \S  \ref{sec:DGP} we showed how
consistency relations, obtained from the evolution and constraint
equations obeyed by perturbation variables, help to distinguish between DE and
MG. The violation of the consistency conditions that can occur with 
sufficient observables (see \S \ref{sec:threevariables}) 
would thus imply either the modification of gravity or 
a coupling of the ``dark energy'' with matter in a GR scenario.

There are several assumptions and caveats in this study. We study MG 
theories with scalar perturbations to the metric; our formalism
does not apply to theories with additional vector or tensor degrees of
freedom. While describing the quasilinear regime of large-scale structure
(where planned observations have the best signal-to-noise), one needs to be
aware that it is not clear how to obtain nonlinear predictions
of some MG theories. We have chosen to work in Fourier space, where the
description of clustering is simpler, but this means that there is not
always a direct relationship of our effective functions for MG
theories with the real space description of these theories
(e.g. $\tilde{G}_{\rm eff}$ is related to its real space counterpart
in the Poisson equation by a convolution with the density field). And
finally, we have left for future work a detailed study of the accuracy
with which MG tests can be performed by the next generation of surveys. 

\section{Appendix: Perturbation theory in modified gravity}

The fluid equations in the Newtonian regime are given by the
continuity, Euler and Poisson equations. Keeping the nonlinear terms
that have been discarded in the study of linear perturbations in the
rest of the paper, the continuity equations gives: 
\begin{equation}
\dot{\delta} + \theta = -\int \frac{d^3 k_1}{(2\pi)^3}  \frac{{\vec 
k}\cdot{\vec k_1}}{k_1^2} \theta(\vec{k_1})\delta({\vec k}-{\vec k_1})
\label{eqn:continuityNL}
\end{equation}
where the term on the right shows the nonlinear coupling of
modes. Note that the time derivatives are with respect to conformal
time in this Appendix. The Euler equation is
\begin{equation}
\dot{\theta} + H \theta - k^2 \psi = 
-\int \frac{d^3 k_1}{(2\pi)^3}  
\frac{k^2{\vec k_1}\cdot(\vec k- \vec k_1)}{2 k_1^2 |{\vec k_1} - 
{\vec k_2^2}|^2 }
\theta(\vec{k_1})\theta({\vec k}-{\vec k_1})
\label{eqn:eulerNL}
\end{equation}
We neglect pressure and anisotropic stress as the energy density is
taken to be dominated by non-relativistic matter \cite{Jain1994}. 
The Poisson equation is given by Eqn. \ref{eqn:MG3} and
supplemented by the relation between $\psi$ and $\phi$ given by
Eqn. \ref{eqn:MG4}. Using these equations we can substitute for
$\psi$ in the Euler equation to get 
\begin{eqnarray}
\dot{\theta} &+&  H \theta + \frac{8 \pi \tilde{G}_{\rm eff}}{(1+\eta)} 
\bar{\rho}_{\rm MG}a^2\delta \nonumber \\
&=& \, -\int \frac{d^3 k_1}{(2\pi)^3}  
\frac{k^2{\vec k_1}\cdot({\vec k- \vec k_1})}{2 k_1^2 |{\vec k_1} - 
{\vec k_2^2}|^2 }
\theta(\vec{k_1})\theta({\vec k}- {\vec k_1})
\label{eqn:eulerNL2}
\end{eqnarray}
Eqns. \ref{eqn:continuityNL} and \ref{eqn:eulerNL2} are 
two equations for the two variables $\delta$ and
$\theta$. They constitute a fully nonlinear
description of MG theories and can be solved once $\eta$ and $\tilde{G}_{\rm
  eff}$ are specified. An important caveat is that they may
nevertheless be invalid for particular theories, for example if the
superposition principle is violated. 

Next we consider perturbative expansions for the density field and the
resulting behavior of the power spectrum and bispectrum. Let $\delta =
\delta_1 + \delta_2 +...$ where 
$\delta_2 \sim O(\delta_1^2)$. 
Higher order effects due to gravitational dynamics become detectable on
10s of Mpc at low redshift. While this is strictly true only for
general relativity, any MG theory that is close enough to GR to fit
observations can also be expected to have this feature. In the
quasilinear regime, 
i.e. on length scales between $\sim$10-100 Mpc, mode coupling effects can be
calculated using perturbation theory. 
For MG, let us simplify the notation by introducing the function: 
\begin{equation}
\zeta_{MG}(k,t) = \frac{8 \pi \tilde{G}_{\rm eff}}{(1+\eta)} ,  
\label{eqn:MGfunction}
\end{equation}
which is simply $4\pi G$ in GR but can vary with time and scale in MG
theories.  
The evolution of the linear growth factor is given by substituting for
$\psi$ in Eqn. \ref{eqn:growth} to get 
\begin{equation}
\ddot{\delta_1}+ H \dot{\delta_1} - \zeta_{MG} \bar{\rho}_{MG}a^2
\delta_1 = 0 . 
\label{eqn:growth}
\end{equation}
In GR, the relation of $\psi$ to $\delta$ is given by the Poisson
equation with constant $G$. In MG, this relation involves both $\eta$
and $\tilde{G}_{\rm eff}$. If either of these functions have a dependence on
$k$ or $z$, then the solution for the growth factor changes. 
The linear solutions for $\psi$ and $\psi+\phi$ are then simply obtained using 
Eqns. \ref{eqn:MG3} and \ref{eqn:MG4}. 

We show below that in addition 
the second order solution has a functional dependence on 
$\tilde{G}_{\rm eff}$ and $\eta$ that can differ. 
Thus potentially distinct signatures of the scale and time dependence of 
$\tilde{G}_{\rm eff}(k,z)$ can be inferred from higher order terms. 
These rely either on features in $k$ and $t$ in 
measurements of $P_{\psi+\phi}$ and $P_\delta$, or
on the three-point functions, which even at 
a single redshift can have distinct signatures of MG
\cite{Bernardeau2004}.  Quasilinear signatures due to $\eta(k,z)$ can also 
be detected via second order terms in the redshift distortion
relations for the power spectrum and bispectrum. Our discussion
generalizes that of \cite{Yukawa}
who examined a Yukawa-like modification of the Newtonian potential. 

\subsection{Second order solution}
From a perturbative treatment of Eqns. \ref{eqn:continuityNL} 
and \ref{eqn:eulerNL2} the second order term for the
growth of the density field is given by 
\begin{eqnarray}
&&\ddot{\delta_2} + H\dot{\delta_2} - \bar{\rho}_{\rm MG}a^2
\zeta_{MG} \delta_2 = \nonumber \\
&& H I_1[\dot{\delta_1}, \delta_1] 
+ I_2[\dot{\delta_1},\dot{\delta_1}] + \dot{I_1}[\dot{\delta_1},  \delta_1] , 
\label{eqn:D2}
\end{eqnarray}
where $I_1$ and $I_2$ denote convolution like integrals of the two
arguments shown, given by the right-hand side of equations 
\ref{eqn:continuityNL} and \ref{eqn:eulerNL2} as follows
\begin{equation}
I_1[\dot{\delta_1},\delta_1](\vec{k}) = \int \frac{d^3 k_1}{(2\pi)^3}
  \frac{{\vec k}\cdot{\vec
    k_1}}{k_1^2} \dot{\delta_1}(\vec{k_1})\delta_1({\vec k}-{\vec k_1})
\end{equation}
and
\begin{equation}
I_2[\dot{\delta_1},\dot{\delta_1}](\vec{k}) = 
\int \frac{d^3 k_1}{(2\pi)^3}  
\frac{k^2{\vec k_1}\cdot(\vec k- \vec k_1)}{2 k_1^2 |{\vec k_1} - 
{\vec k_2^2}|^2 }
\dot{\delta_1}(\vec{k_1})\dot{\delta_1}({\vec k}-{\vec k_1}) .
\end{equation}
Finally, the last term in Eqn. \ref{eqn:D2} is simply 
$\dot{I_1}[\dot{\delta_1},  \delta_1] = I_1[\ddot{\delta_1}, \delta_1] +
I_1[\dot{\delta_1},\dot{\delta_1}]$. Note that by continuing the iteration 
higher order solutions can be obtained. 

From the above equations it follows  that 
if $\zeta_{MG}\equiv \zeta_{MG}(t)$ then $\delta_2$ may be specified by
$k$-integrals over $\delta_1$, so that one may express the functional
relationship  $\delta_2 \equiv \delta_2[\delta_1]$ (where it is
understood that $\delta_2$ at a given wavenumber 
$\vec{k}$ depends on $\delta_1$ at all other wavenumbers). But for the
general case of a MG theory with scale dependent $\zeta_{MG}$, the second
order solution has additional scale and time dependence behavior that
is not determined by the linear solution (owing to the third term on
the left hand side in Eqn. \ref{eqn:D2}). So the
functional relationship must be modified to: 
$\delta_2\equiv \delta_2[\delta_1; \zeta_{MG}]$. This means that 
quasilinear evolution provides an additional signature of MG. 
That is, even if the initial power spectrum is
not fully specified (e.g. if the running of the spectral index is not
well constrained), the comparison of linear and quasilinear growth
rates can reveal the signature of MG. In practice, whether the quasilinear
signature is significant must be determined by computations for
specific models (see \cite{Stabenau2006} for a specific model for
which it is not). Note also that the second order
correction to the density power spectrum also involves the third order
density field as it is given by  $P_2 \sim \langle \delta_2^2\rangle +
\langle \delta_1 \delta_3 \rangle$. The qualitative features we
highlight for $\delta_2$ will also be found in $\delta_3$, which is also
given by iterations of the nonlinear Eqns. \ref{eqn:continuityNL} and
\ref{eqn:eulerNL}. 

We summarize the comparison of linear and second order solutions for
the density for GR versus MG. We have identified the function
$\zeta_{MG}(k,t)$ as containing all the information about MG that
affects density and velocity fields. For the density field the first
and second order solutions can be compared to GR as: 
\begin{itemize}
\item {\it Linear growth in GR:} In smooth dark energy GR models,
  $\delta_1(\vec k,t)$ is a separable  function of scale and time. 
\item  {\it Linear growth in MG:} In MG theories, $\delta_1(\vec k,t)$
  is a separable function of $k$ and $t$ if and only if the MG function 
$\zeta_{MG}$ is independent of scale. 
\item  {\it Second order solution in GR:}
In smooth dark energy GR models, the second order solution
  $\delta_2(\vec k, t)$ is not separable. It is however determined by
  integrals over $\delta_1$. 
\item {\it Second order solution in MG:}
In MG models with $\zeta_{MG}(k,t)$, $\delta_2$ is no longer
  determined solely by $\delta_1$ and contains additional 
 signatures of MG. 
\end{itemize}
Note that for weak lensing measurements, quasilinear corrections are
given by the density times $\tilde{G}_{\rm eff}$ (by substituting higher
order terms into Eqn.\ref{eqn:power4}). So the resulting
signatures can be straightforwardly computed using the higher order
solutions for the density field. 

\subsection{Three-point correlations}
Distinct quasilinear effects are found in three-point correlations
(we will use the Fourier space bispectrum), as it is the lowest order probe
of gravitationally induced non-Gaussianity. The bispectrum for the
density field $B_\delta$ is defined by
\begin{equation}
\langle \delta({\vec k_1}) \delta({\vec k_2})\delta({\vec k_3}) \rangle = 
(2 \pi)^3 \delta_{\rm D}({\vec k_1 + \vec k_2 + \vec k_3}) 
B_\delta(\vec k_1, \vec k_2, \vec k_3).
\label{eqn:bispectrum}
\end{equation}
Since $B_\delta\sim \langle \delta^3\rangle\sim \langle \delta_1^2
\delta_2\rangle$ (using $\langle\delta_1^3\rangle=0$ for an initially
Gaussian density field), 
the second order solution enters at leading order in the
bispectrum. Note also that the wavevector arguments of the bispectrum
form a triangle due to the Dirac delta function on the right-hand
side above. In practice, a very useful measure of non-Gaussianity
is the reduced bispectrum function $Q$, which for the density field 
$\delta$ is given by 
\begin{equation}
Q_\delta
\equiv \frac{B_\delta(\vec k_1, \vec k_2, \vec k_3)}
{P_\delta(k_1)P_\delta(k_2) + P_\delta(k_2)P_\delta(k_3) + 
P_\delta(k_1)P_\delta(k_3) }
\end{equation}
To leading order $Q$ is independent of
the amplitude of the linear 
power spectrum (both numerator and denominator are O($\delta_1^4$), 
see \cite{Bernardeau2004})
and is nearly constant with triangle size in GR. It is however
sensitive to the shape of the triangle. The dependence on size and
shape changes for MG theories and is in principle a probe of
$\zeta_{MG}$.  It is beyond the scope of this paper to
elaborate on the measurement of the bispectrum from galaxy surveys; 
we will instead focus on the prospects for lensing measurements. 
\cite{Shirata2007} have tested Yukawa like modifications of gravity
using $Q$ for the galaxy density measured in real and redshift space. 

The lensing bispectrum contains perhaps the clearest signature of MG.  
It is a projection of the three dimensional bispectrum
\begin{equation}
k^6 B_{\psi+\phi} \sim 
(8\pi \tilde{G}_{\rm eff}a^2{\bar{\rho}_{\rm MG}})^3
\langle\delta^3\rangle 
\simeq (8\pi \tilde{G}_{\rm eff}a^2{\bar{\rho}_{\rm MG}})^3
\langle \delta_1^2 \delta_2 \rangle 
\end{equation}  
Since both $\delta_1$ and $\delta_2$ are 
function of $\zeta_{MG}$, measurements of
$B_{\psi+\phi}$ are sensitive to $\tilde{G}_{\rm eff}$ and $\zeta_{MG}$
separately. 

The reduced lensing bispectrum in a MG theory can be expressed in
terms of the density power spectrum and bispectrum as:  
\begin{equation}
Q_{\psi+\phi} \propto
\frac{\tilde{G}_{\rm eff}(k_1)\tilde{G}_{\rm eff}(k_2)\tilde{G}_{\rm
  eff}(k_3) B_\delta(\vec k_1, \vec k_2, \vec
  k_3)/{\bar{\rho}_{\rm MG}}}
{k_3^2\tilde{G}_{\rm eff}^2(k_1)P_\delta(k_1)\tilde{G}_{\rm
    eff}^2(k_2)P_\delta(k_2)/k_1^2 k_2^2
  + {\rm sym}...} 
\end{equation}

For equilateral triangles, $Q$ in MG theories is simpler
since the $G_{\rm eff}$ factors in all the terms are the same.  
One then has 
\begin{equation}
Q_{\psi+\phi(MG)}
\propto 
\frac{ k^2 Q_{\delta(MG)}} 
{\tilde{G}_{\rm eff}\ {\bar{\rho}_{\rm MG}}}
\end{equation}
The ratio of $Q$ for MG versus GR for equilateral triangles is given by 
\begin{equation}
\frac{ Q_{\psi+\phi(MG)}} {Q_{\psi+\phi(GR)}}
\propto 
\frac{ Q_{\delta(MG)}} {Q_{\delta(GR)}}
\frac{G\ {\bar{\rho}_{\rm GR}}} {\tilde{G}_{\rm eff}\ {\bar{\rho}_{\rm MG}}}
\end{equation}
Note that $Q_\delta$ itself depends on $\zeta_{MG}$. 
Bernardeau \cite{Bernardeau2004} shows that with $\eta=1$ but a scale dependent
$\tilde{G}_{\rm eff}$, $Q_\delta$ for given initial power spectrum is
relatively insensitive to $\zeta_{MG}$. If that holds for generic
intial power spectra and gravity models,
it would imply that $Q_{\psi+\phi(MG)}$ probes 
$\tilde{G}_{\rm eff}$ for models with $\eta=1$. 

In general a measurement of the lensing power spectrum and reduced
bispectrum (roughly speaking, of $P_{\psi+\phi}$ and $Q_{\psi+\phi}$) is
sufficient to measure departures from GR. There are three underlying functions 
($P_\delta$, $\zeta_{MG}$ and $\tilde{G}_{\rm  eff}$) to be
determined. For given $k$ and source redshift, we have measurements of
$P$ and of $Q$ as a function of triangle shape. Thus while the
equilateral triangles may be regarded as sensitive primarily 
to $\tilde{G}_{\rm  eff}$,
elongated triangles will be sensitive to $\zeta_{MG}$, and
therefore to $\eta$. 
In practice one must take account of the fact
that the bispectrum has lower signal-to-noise than the power spectrum
on quasilinear scales \cite{Takada2004}, so one must fit for the desired 
information from all triangle configurations and sizes to constrain
the MG functions. 

To summarize this section, quasilinear effects thus offer two signatures of MG. 
\begin{itemize}
\item
A scale and time dependent feature on quasilinear scales 
in the power spectrum that depends on $\eta$ and $\tilde{G}_{\rm eff}$. This
enters through the second order contribution to the power spectrum.
\item
Signatures in the bispectrum: additional signatures of modified
gravity are present in three-point correlations of the density and
potential fields. Independent of the shape and amplitude of the power
spectrum, the dependence of the reduced bispectrum $Q$ on triangle
size and shape is a useful test of MG. The reduced lensing bispectrum for
example has a strong dependence on $\tilde{G}_{\rm eff}$. 
\end{itemize}
We have not considered here whether a clustered DE model can
mimic both these signatures. It would be of interest to carry out the
second order calculations for a set of MG models and compare predicted
deviations with observational error bars. 

\bigskip

{\it Acknowledgments:}
We are grateful to Jacek Guzik, Eric Linder and Wayne Hu for helpful
discussions and comments on an early draft. We thank Francis
Bernardeau, Raul Jimenez, Matt Martino, Roman Scoccimarro, Ravi Sheth, Fritz
Stabenau, Masahiro Takada, Jean-Philippe Uzan and Licia Verde 
for stimulating discussions. PJZ is supported by the one-hundred talents
program of the Chinese Academy of Science (CAS), the National Science
Foundation of China  grant 10533030 and CAS grant KJCX3-SYW-N2. 
BJ is supported in part by NSF grant AST-0607667, the
Department of Energy and the Research Corporation.

\end{document}